\documentclass[aps,prm,twocolumn]{revtex4}

\usepackage{graphicx}
\usepackage{dcolumn}
\usepackage{amsmath}
\usepackage{amssymb}
\usepackage{wasysym}
\usepackage{color}

\usepackage{mathtools}

\begin{document}

\title{Adaptation of Wallace's approach to the specific heat of elemental solids with significant intrinsic anharmonicity, particularly the light actinide metals
}

\author{Christopher A. Mizzi, W. Adam Phelan, Matthew S. Cook, Greta L. Chappell, Paul H. Tobash, David C. Arellano,  Derek V. Prada, Boris Maiorov, and Neil~Harrison}

\affiliation{
Los~Alamos~National Laboratory, Los~Alamos,~NM~87545
}
\date{\today}













\section*{}
\noindent 

\begin{abstract}\noindent
The quasiharmonic approximation is the most common method for modeling the specific heat of solids; however, it fails to capture the effects of intrinsic anharmonicity. In this study, we introduce the ``elastic softening approximation,'' an alternative approach to modeling intrinsic anharmonic effects on thermodynamic quantities, {which is grounded in Wallace's thermodynamic framework} that tracks entropy changes resulting from the continuous \textcolor{black}{change (e.g.}, softening) of phonons as a function of temperature. A key finding of our study is a direct correlation between Poisson’s ratio and the differential rate of phonon softening at finite frequencies, compared to lower frequencies relevant to elastic moduli measurements. We observe that elemental solids such as $\alpha$-Be, diamond, Al, Cu, In, W, Au, and Pb, which span a wide range of Poisson’s ratios and exhibit varying degrees of intrinsic anharmonicity, consistently follow this trend. When applied to $\alpha$-U, $\alpha$-Pu, and $\delta$-Pu, our method reveals unusually large anharmonic phonon contributions at elevated temperatures across all three light actinide metals. These findings are attributed to the unique combination of enhanced covalency and softer elastic moduli inherent in the actinides, potentially influenced by their $5f$-electron bonding. 
\end{abstract}

\maketitle 

\section{Introduction}

To understand how materials behave under extreme conditions, it is crucial to know how entropy changes with temperature~\cite{blanco2004,lomonosov2007,sjostrom2016,mchardy2018}. One method to assess the predictability of phonons involves the quasiharmonic approximation~\cite{fultz2010}, which extends the harmonic model to consider the impact of thermal expansion and contraction against a bulk modulus. The quasiharmonic approach models the specific heat of phonons, $C_{p,{\rm ph}}(T)$, as a function of temperature at constant pressure, starting with the specific heat at constant volume, $C_{v,{\rm ph}}(T)$, calculated from the phonon density of states as in a harmonic crystal~\cite{ashcroft1976}. It then incorporates a term that factors in thermal expansivity and the isothermal bulk modulus~\cite{fultz2010}. Such an approach proves to be especially useful in systems exhibiting small shifts in phonon frequencies with temperature attributable to intrinsic anharmonicity, such as solid Al~\cite{kresch2008}.

Problems arise with the quasiharmonic approximation when intrinsic anharmonicity effects become significant, as these effects are neglected in this approach. By ``intrinsic anharmonicity,'' we refer to systems where notable changes in phonon frequencies occur under conditions of near constant volume~\cite{fultz2010, jacobs2005}. While intrinsic anharmonic effects become important in most solids at sufficiently high temperatures, it dominates over a larger temperature range in materials that simultaneously exhibit \textcolor{black}{phonon softening with increasing temperature and a small or negative thermal expansion}~\cite{dove2016}.  Such effects are especially pronounced in the actinide metal $\delta$-Pu~\cite{wallace1998, lawson2019,bottin2024}. In addition to exhibiting a negative thermal expansivity at high temperatures~\cite{lawson2006,jette1954,schonfeld1996}---a rare trait for an elemental solid---it also exhibits an anomalously large phonon softening with increasing temperature, as documented by elastic moduli measurements~\cite{suzuki2011} and temperature-dependent neutron scattering measurements of the phonon density of states~\cite{mcqueeney2004}.

\begin{figure*}
\begin{center}
\includegraphics[width=0.95\linewidth]{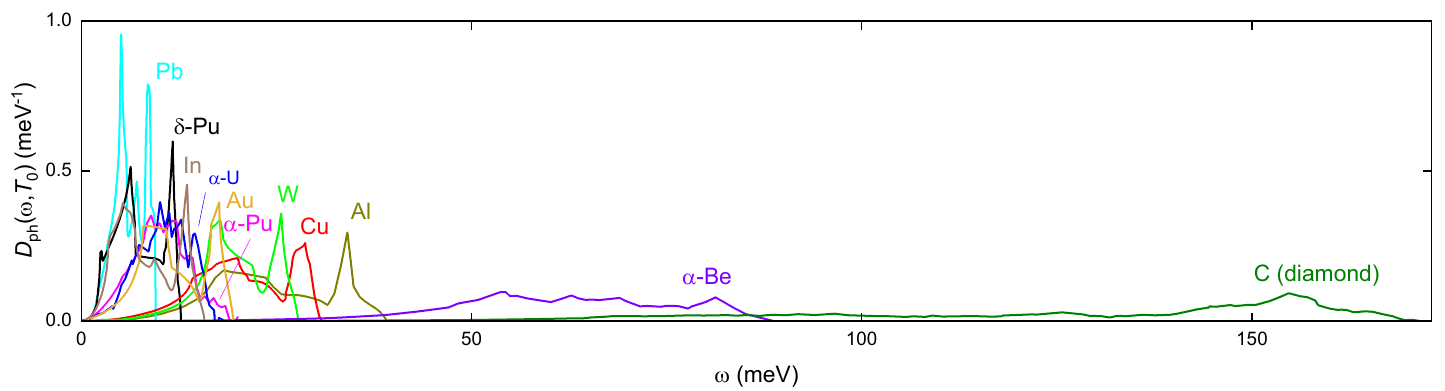}
\vspace{-4mm}
\textsf{\caption{
The phonon density of states, $D_{\rm ph}(\omega,T_0)$, measured at temperature $T_0$ for the elemental solids $\alpha$-Be~\cite{schmunk1966,wormald2017}, diamond~\cite{bosak2005}, Al~\cite{walker1956}, Cu~\cite{nilsson1973}, In~\cite{schober1981}, W~\cite{schober1981}, Au~\cite{lynn1973}, Pb~\cite{gilat1965}, $\alpha$-U~\cite{soderlind2021}, $\alpha$-Pu~\cite{manley2009}, and $\delta$-Pu~\cite{wong2005}. Values for $T_0$ are given in Table~\ref{table2}.
}
\vspace{-4mm}
\label{phononDOSfig}}
\end{center}
\end{figure*}

In this paper, we introduce an alternative to the quasiharmonic approximation for modeling the specific heat of elemental solids at high temperatures, which we term the `elastic softening approximation.’ Based on Wallace's approach~\cite{wallace1972}, this method is designed to address the limitations of the quasiharmonic approximation and considers the effect of a continuous softening of  phonon frequencies on the entropy as a function of temperature~\cite{wallace1972,bryan2019,jacobs2005}. \textcolor{black}{Rather than tracking the entire phonon spectrum as a function of temperature (the most accurate approach to modeling thermodynamic quantities~\cite{allen2015}), our method uses elastic moduli measurements, making it significantly simpler to implement. Crucially, we find that it still captures important intrinsic anharmonic phonon effects arising from frequency changes \emph{not} driven by thermal expansion, as observed in $\delta$-Pu~\cite{mcqueeney2004,suzuki2011}.}

We first test our approach on a variety of regular elemental solids with different crystal \textcolor{black}{and electronic} structures, namely $\alpha$-Be, diamond (C), Al, Cu, In, W, Au, and Pb. To fully capture the phonon contribution to specific heat at high temperatures, we introduce what we refer to as a differential softening parameter, $\eta$. This parameter accounts for the differences in the rate of phonon softening at finite frequencies compared to the vanishing frequency limit, which is relevant to elastic moduli measurements. In the few systems where the phonon density of states has been measured at multiple temperatures~\cite{kresch2008,larose1976, semenov2014,manley2001,mcqueeney2004}, we find that $\eta$ accurately predicts the softening of prominent features in the phonon density of states.

We also find that $\eta$ is closely correlated with Poisson’s ratio~\cite{koster1961,ledbetter2008,greaves2019}, suggesting a link between $\eta$ and bonding. A larger $\eta$ (indicating increased phonon softening at finite frequencies relative to ultrasound measurements) is linked to brittleness and more covalent bonding character. Conversely, a smaller $\eta$ (indicating suppressed softening at finite frequencies) is associated with ductility and more metallic bonding character.

We observe that the actinide metals $\alpha$-U, $\alpha$-Pu, and $\delta$-Pu follow the same trend as other solids, once we account for their anomalous electronic contribution~\cite{harrison2023} to the specific heat at low temperatures. However, because the phonon softening in these actinides is much more dramatic than in other elemental solids, their intrinsic anharmonic contribution to the total specific heat is anomalously large, especially in $\delta$-Pu~\cite{oetting1983,adams1983,wallace1998,dai2003,robert2003,migliori2006}.


\section{Approximations to the Specific Heat}

The phonon contribution to the specific heat \(C_{{V,p},{\rm ph}}(T)\) is related to the phononic entropy $S_{\rm ph}(T)$ via~\cite{cohen1996} 
\begin{equation}\label{phononspecificheat}
C_{{V,p},{\rm ph}}(T)=T~\frac{\partial S_{\rm ph}(T)}{\partial T}\bigg|_{V,p}~,
\end{equation}
where $T$ is temperature and the subscripts \(V\) and \(p\) refer to constant volume and constant pressure conditions, respectively. 

While it is sometimes possible to achieve exceptionally good fits to the specific heat and other thermodynamic quantities using methods based on the Debye or Einstein models~\cite{ashcroft1976}, particularly for modeling the equation of state~\cite{kozyrev2022i,holzapfel2001,lomonosov2007,shang2010}, such methods often require multiple adjustable parameters. This reliance on adjustable parameters can obscure the underlying physics we seek to investigate. To gain deeper insight into the limitations or advantages of a given approximation to the specific heat in relation to anharmonic phonon effects, we desire to utilize the measured phonon density of states, as depicted in Fig.~\ref{phononDOSfig}~\cite{schmunk1966,wormald2017, bosak2005, walker1956, nilsson1973, schober1981, lynn1973, gilat1965, soderlind2021, manley2009, wong2005}, to determine the phonon contribution to the specific heat.

\subsection{Harmonic and Quasiharmonic Approximations to Phononic Entropy}

The harmonic approximation (HA) models the phonon contribution to the specific heat at constant volume, $C_{V,\mathrm{ph}}^{\mathrm{HA}}(T)$, by assuming that the phonon frequencies, $\omega$, have no temperature dependence. Under this assumption, the phonon density of states, $D_{\mathrm{ph}}$, is temperature independent [\textit{i.e.,} $D_{\mathrm{ph}}(\omega,T) = D_{\mathrm{ph}}(\omega,T_0)$ where $T_0$ denotes the temperature at which the phonon density of states is measured, often around 300~K (see Table~\ref{table2})]. Temperature is assumed only to affect the phonon contribution to entropy via thermal occupancy. The phononic entropy can then be calculated from the phonon density of states using~\cite{fultz2010,bryan2019} 
\begin{eqnarray}\label{phononenytropy}
S^{\rm HA}_{\rm ph}(T)=R\int_0^\infty D_{\rm ph}(\omega,T_0)\hspace{3.5cm}\nonumber\\\times~ \big[(n_{\rm BE}+1)\ln(n_{\rm BE}+1)-n_{\rm BE}\ln(n_{\rm BE})\big]{\rm d}\omega,
\end{eqnarray}
where \[n_{\rm BE}=\big(e^{\frac{\hbar\omega}{k_{\rm B}T}}-1\big)^{-1}\] is the Bose-Einstein distribution function, and \(R=N_{\rm A}k_{\rm B}\) is the gas constant. Within this approximation, $C_{V,\mathrm{ph}}^{\mathrm{HA}}(T)$ is found by performing the integration in Equation~\ref{phononenytropy} using $D_{\mathrm{ph}}(\omega,T_0)$ and substituting this phonon contribution to entropy into Equation~\ref{phononspecificheat}.

The quasiharmonic approximation (QHA) improves upon the harmonic approximation by considering the effects of a temperature-dependent volume on the phonon contribution to the heat capacity. In quasiharmonic approximation treatments, the lattice is modeled as uniformly expanding and contracting against a bulk modulus~\cite{fultz2010}. The phonon contribution to the heat capacity at constant pressure within the quasiharmonic approximation, $C^{\rm QHA}_{p,{\rm ph}}(T)$, is then estimated as the sum of the harmonic approximation contribution at constant volume and a term describing the effects of the temperature-dependent volume changes,
\textcolor{black}{\begin{equation}\label{quasiharmonic0}
C_{p,\mathrm{ph}}^{\mathrm{QHA}}(T) \approx C_{V,\mathrm{ph}}^{\mathrm{HA}}(T) + T\;\frac{\partial V}{\partial T}\bigg|_p\;\frac{\partial S}{\partial V}\bigg|_T.
\end{equation}
From Maxwell's relations, one obtains:}
\begin{equation}\label{quasiharmonic}
C_{p,\mathrm{ph}}^{\mathrm{QHA}}(T) \approx C_{V,\mathrm{ph}}^{\mathrm{HA}}(T) + V(T)\alpha_V^2(T)B_T(T)T, 
\end{equation}
where $V(T)$ is the molar volume, $\alpha_V(T)$ the thermal expansivity, and $B_T(T)$ the isothermal bulk modulus  (see Appendix A). The temperature-dependences of these quantities are nearly always neglected~\cite{fultz2010}.

\begin{table*} \textcolor{black}{
\begin{tabular}{||c|c|c|c|c|c|c|c|c||} 
 \hline
 element & $T_0$~(K) & $V(T_0)$~(cm$^3$mol$^{-1}$) & $B_T(GPa)$&$\Gamma_{\rm el}$ (mJmol$^{-1}$K$^{-2}$)&$\gamma_{\rm el}$&$\eta$&$\sigma\Delta C_p^{\rm AN}$ (Jmol$^{-1}$K$^{-1}$)&$\nu(T\rightarrow0)$ \\ [0.5ex] 
 \hline\hline
$\alpha$-Be & 300&4.851 & 107.4 &0.17&-&1.83~$\pm$~0.11&0.15&0.042 \\ 
 \hline
C (diamond) & 10& 3.416 &444.2 &0.00&0&1.44~$\pm$~0.68&0.51&0.070\\
 \hline
Al & 300& 10.002 & 72.7 &1.26&1.62&0.62~$\pm$~0.05&0.20&0.335 \\
 \hline
Cu & 80& 7.044 & 140.3 &0.69&0.92&0.63~$\pm$~0.06&0.15&0.340 \\
 \hline
 In & 77& 15.469 & 45.7 &1.66&3&0.26~$\pm$~0.01&0.06&0.420 \\
 \hline
W & 298& 9.551 & 307.2 &1.01&0.3&1.15~$\pm$~0.07&0.16&0.279 \\
 \hline
Au & 300& 10.216 & 172.3 &0.69&1.6&0.39~$\pm$~0.07&0.35&0.424 \\
 \hline
Pb & 300& 18.273 & 41.6 &2.93&1.7&0.30~$\pm$~0.02&0.08&0.393 \\ 
\hline
$\alpha$-U & 300& 12.492 & 121.6 &9.14&-&1.23~$\pm$~0.20&1.17&0.212 \\ 
\hline
$\alpha$-Pu & 300& 12.307 & 48.8 &17&-&1.45~$\pm$~0.20 &0.50&0.179\\ 
\hline
$\delta$-Pu & 300& 14.892 & 30.6 &64&-&0.97~$\pm$~0.12&0.58&0.273 \\ 
 \hline
\end{tabular}}
\caption{\label{table2} Values of temperature $T_0$ for each element at which the phonon density of states was measured (see Fig.~\ref{phononDOSfig}), along with the values of molar volume $V(T_0)$, and the isothermal bulk modulus $B_T(T_0)$ at this temperature. The table also includes the Sommerfeld coefficient for each elemental solid, \textcolor{black}{the electronic Gr\"{u}neisen parameter}, the obtained values of $\eta$, \textcolor{black}{the estimated error in anharmonicity~$\sigma\Delta C_p^{\rm AN}$ in Fig.~\ref{comparisonfig}b averaged over temperature}, and the low-temperature Poisson's ratio $\nu$ (determined from the data in Fig.~\ref{shear}). For $\delta$-Pu, $V(T_0)$ corresponds to samples in which the $\delta$ phase was stabilized by substituting 2 atomic percent of the Pu sites with Ga~\cite{lawson2006}. \textcolor{black}{Source references are provided in the text}.}
\end{table*}

\textcolor{black}{It should be noted that the second term on the right-hand side of Equation~\ref{quasiharmonic} derives from the standard thermodynamic identity~\cite{fultz2010}:
\begin{equation}\label{cpminuscv}
C_{p}(T) - C_{V}(T) = V(T)\,\alpha_V^2(T)\,B_T(T)\,T.
\end{equation}
By using experimental values for $\alpha_V(T)$ and $B_T(T)$, one implicitly includes a small electronic contribution, which arises from the volume dependence of the electronic entropy~\cite{varley1956}; hence, the use of `$\approx$' in Equation~\ref{quasiharmonic}. In conventional metals, this volume-dependent electronic contribution is very small and is rarely mentioned in discussions of the quasiharmonic approximation. In insulators, such an electronic contribution is absent. We return to a discussion of the volume-dependent electronic term in Section IID.}


\subsection{Elastic Softening Approximation to Phononic Entropy}

\textcolor{black}{The primary limitation of the quasiharmonic approximation is that it attributes the temperature dependence of the entropy at constant volume only to changes in thermal occupancy and omits any explicit temperature dependence of the phonon frequencies themselves. To move beyond the quasiharmonic approximation, one must recognize that the total change in frequency with temperature can be decomposed into two parts:
\begin{equation}\label{twocomponents}
\mathrm{d}\omega(V,T)
\;=\;
\frac{\partial \omega}{\partial V}\bigg|_{T}\mathrm{d}V
\;+\;
\frac{\partial \omega}{\partial T}\bigg|_{V}\mathrm{d}T.
\end{equation}
The quasiharmonic approximation incorporates only the first component, 
${\partial \omega}/{\partial V}|_T$~\cite{fultz2010}, and thus neglects ${\partial \omega}/{\partial T}|_V$, the term responsible for intrinsic anharmonic effects~\cite{dove2016,fultz2010,jacobs2005,holzapfel2005}.} Intrinsic anharmonicity arises from a number of sources, including phonon-phonon and electron-phonon interactions~\cite{fultz2010}. It can be substantial in many systems at high temperatures~\cite{wallace1997} and pressures~\cite{jacobs2005}, and is expected to be dominant in systems with near zero thermal expansivity~\cite{dove2016}. However, the treatment of intrinsic anharmonicity is still an open question.

\textcolor{black}{While early studies of anharmonicity had identified the important role of temperature-dependent frequency shifts at constant volume in thermodynamic quantities~\cite{varley1956,barron1965}, a particularly elegant approach to treating intrinsic anharmonicity was that later introduced by Wallace~\cite{wallace1972}. He showed that, to first order, anharmonic phonon contributions to the entropy can be incorporated by substituting experimental temperature-dependent phonon frequencies into the the density of states in Equation~\ref{phononenytropy}~\cite{jacobs2005}.} Therefore, by replacing \(D_{\rm ph}(\omega,T_0)\) with \(D_{\rm ph}(\omega,T)\) and performing the integration in Equation~\ref{phononenytropy}, in principle, one can calculate \(C_{{p},{\rm ph}}(T)\) using Equation~\ref{phononspecificheat} over any desired parameter space provided experimental values of \(D_{\rm ph}(\omega,T)\) are known. Note, we focus on the heat capacity at constant pressure because this is often the most experimentally accessible quantity. In practice, widespread application of Wallace's method is constrained by the limited availability of \(D_{\rm ph}\big(\omega,T)\) data, often only available for a select range of temperatures and materials. 

Without measurements of the entire density of states over a wide temperature range, the challenge is then to figure out ways to approximate the temperature dependence of the phonon frequencies. To this end, we propose our elastic softening approximation (ESA), which assumes that the temperature-dependence of phonon frequencies can be quantified by the factor \(\beta_p(T)\). This factor is defined such that \(\beta_p(T)=\omega(T)/\omega(T_0) \), where $\omega$ refers to a reference frequency in the phonon spectrum, with \(\beta_p(T_0) = 1\) at the reference temperature \(T_0\). In this approximation, the phonon density of states undergoes a continuous temperature-dependent rescaling relative to the phonon density of states at $T_0$, as represented by the equation:
\begin{equation}\label{phononDOS}
    D_{\mathrm{ph}}(\omega,T) \approx \beta_p^{-1}(T) D_{\mathrm{ph}}\big(\omega\beta_p^{-1}(T),T_0\big),
\end{equation}
resulting in a shift of the phonon density of states to lower frequencies at higher temperatures, while maintaining the total number of phonon states.

The impact of temperature-dependent phonon frequency changes on entropy is a critical initial consideration prior to deriving other thermodynamic quantities, such as the specific heat~\cite{wallace1972,bryan2019,jacobs2005}. We do this by substituting $D_{\mathrm{ph}}(\omega,T)$ in Equation~\ref{phononDOS} for $D_{\mathrm{ph}}(\omega,T_0)$ into Equation~\ref{phononenytropy} and then determining the specific heat using Equation~\ref{phononspecificheat}. 

This approach is consistent with Wallace's approach as shown by introducing a variable change: \(\omega^\prime = \beta_p^{-1}(T)\omega\). This leads to an altered expression for the entropy:
\begin{align}\label{phononenytropySA}
S^{\rm ESA}_{{\rm ph}}(T)=R\int_0^\infty D_{\rm ph}(\omega^\prime,T_0)\hspace{3.5cm}\nonumber\\~\times~ \big[(n^\prime_{\rm BE}+1)\ln(n^\prime_{\rm BE}+1)-n^\prime_{\rm BE}\ln(n^\prime_{\rm BE})\big]{\rm d}\omega^\prime,
\end{align}
resembling Equation~\ref{phononenytropy}, but where the Bose-Einstein distribution function is modified to \[n^\prime_{\rm BE} = \big(e^{\frac{\hbar\omega^\prime \beta_p(T)}{k_{\rm B} T}} - 1\big)^{-1}.\] At \(T = T_0\), where \(\beta_p(T=T_0) = 1\), the entropy described by Equation~\ref{phononenytropySA} aligns with the harmonic approximation at temperature $T_0$. We show consistency with Wallace's approach explicitly by considering the frequency shift as a small perturbation: \(\delta\omega \approx \omega(\beta_p(T) - 1)\). Upon substituting $\beta_p(T)\approx1+\delta\omega/\omega$ into Equation~(\ref{phononenytropySA}), we obtain:
\begin{align}\label{wallacetheorem}
S^{\rm ESA}_{{\rm ph}}(T)\approx S^{\rm HA}_{{\rm ph}}(T)\hspace{3.5cm}\nonumber\\~-\delta\omega\times N_{\rm A}\int_0^\infty D_{\rm ph}(\omega,T_0)\frac{\partial n_{\rm BE}}{\partial T}{\rm d}\omega,
\end{align}
which aligns with Wallace's expression for the anharmonic phonon contribution~\cite{wallace1972,jacobs2005}. The negative \(\delta\omega\) at high temperatures indicates that phonon softening invariably results in an increase in entropy. 

\textcolor{black}{Because the exponent in the Bose-Einstein distribution function is the only part of Equation~\ref{phononenytropySA} that is modified in the elastic softening approximation, the entropy remains as straightforward to compute by integration as it is in the harmonic approximation. Moreover, it is more accurate for large changes in frequency than the perturbative approximation given by Equation~\ref{wallacetheorem}. 
Consequently, we use Equation~\ref{phononenytropySA} throughout the paper to calculate $S^{\rm ESA}_{\mathrm{ph}}(T)$.}

\textcolor{black}{Interestingly}, the modified form of the Bose-Einstein distribution function in Equation~\ref{phononenytropySA} implies that the effect of $\beta_p(T)$ in our elastic softening approximation is to \textcolor{black}{effectively} rescale $S^{\rm HA}_{\rm ph}(T)$ in temperature, so that 
\begin{equation}\label{simpleentropyscaling}
S^{\rm ESA}_{{\rm ph}}(T)=S^{\rm HA}_{{\rm ph}}\Big(\frac{T}{\beta_p(T)}\Big)\nonumber.
\end{equation}
\textcolor{black}{
In the high-temperature regime, where $C_{V,\mathrm{ph}}^{\mathrm{HA}}(T)\to 3R$ (i.e.\ the Dulong--Petit limit) and $S_{\mathrm{ph}}^{\mathrm{HA}}(T)$ varies logarithmically with $T$, the elastic softening approximation yields
\begin{equation}\label{specificheatAA}
C_{p,\mathrm{ph}}^{\mathrm{ESA}}(T)
\;\to\;
3R\Bigl(1 - T\,\frac{\partial \ln \beta_p(T)}{\partial T}\Big|_p\Bigr),
\end{equation}
where
\begin{equation}\label{betaderivative}
\frac{\partial \ln \beta_p(T)}{\partial T}\Big|_{p}
\;=\;
\frac{\partial \ln \beta_p(T)}{\partial T}\Big|_{V}
\;+\;
V(T)\,\alpha_V(T)\,\frac{\partial \ln \beta_p(T)}{\partial V}\Big|_{T}.
\end{equation}
The first term on the right-hand side of Equation \ref{betaderivative} is equivalent to the standard definition of {intrinsic} anharmonicity (see Section III.D). Meanwhile, the second term is expected to be captured by the quasiharmonic approximation.}

\subsection{Estimating Phonon Frequency Changes from Elastic Moduli}

Within the elastic softening approximation outlined above, the treatment of intrinsic anharmonicity amounts to finding a suitable method to quantify $\beta_p(T)$. We accomplish this by initially considering the elastic moduli, which are second derivatives of free energy with respect to strain~\cite{luthi2005}. Polycrystals with equiaxial distributions of grain orientation (the scope of this work) possess two independent elastic moduli, taken here to be the adiabatic longitudinal $L_S(T)$ and shear $G_S(T)$ elastic moduli. The former is related to the bulk and shear moduli via $L_S(T)=B_S(T)+4G_S(T)/3$. Each of these isotropic moduli possess distinct temperature-dependences which, in the absence of phase transitions, describe the evolution of phonon frequencies with temperature in the limit $\omega\rightarrow0$~\cite{luthi2005}.

More explicitly, the above elastic moduli can be recast as sound velocities which are the gradient of the acoustic branches of the phonon dispersion in the long-wavelength limit. The two elastic moduli considered in this work are related to the longitudinal, $c_{\rm long}(T)$, and transverse, $c_{\rm tran}(T)$, sound velocities according to $c_{\rm long}(T) = \sqrt{L_S(T) / \rho(T)}$ and $c_{\rm tran}(T) = \sqrt{G_S(T) / \rho(T)}$, where $\rho(T)$ is the temperature-dependent density. Relative changes in these sound velocities with temperature from some reference temperature $T_0$ are then given by:

\begin{equation}\label{soundmodulusscalling1}
\frac{c_{\rm long}(T)}{c_{{\rm long}}(T_0)}=\sqrt{\frac{L_S(T)\rho(T_0)}{L_S(T_0)\rho(T)}},
\end{equation}
\begin{equation}\label{soundmodulusscalling2}
\frac{c_{\rm tran}(T)}{c_{{\rm tran}}(T_0)}=\sqrt{\frac{G_S(T)\rho(T_0)}{G_S(T_0)\rho(T)}}.
\end{equation}

Combining the definition of sound velocities in the $\omega\rightarrow0$ limit ($\omega = c k$) with the definition of the elastic softening parameter introduced above ($\beta_p(T) = \omega(T) / \omega(T_0)$), the relative change in sound velocities with temperature will scale as:
\begin{equation}\label{soundfrequencyscaling1}
\frac{c_{\rm long}(T)}{c_{{\rm long}}(T_0)}=\beta_{p,{\rm long}}(T)\frac{k(T_0)}{k(T)},
\end{equation}
\begin{equation}\label{soundfrequencyscaling2}
\frac{c_{\rm tran}(T)}{c_{{\rm tran}}(T_0)}=\beta_{p,{\rm tran}}(T)\frac{k(T_0)}{k(T)}.
\end{equation}
Here, ${k(T_0)}/{k(T)}$ adjusts for the temperature dependence of the reciprocal lattice size, with $k(T)\propto1/\sqrt[3]{V(T)}$.

Combining Equations~\ref{soundmodulusscalling1} and
~\ref{soundfrequencyscaling1}, and also Equations~\ref{soundmodulusscalling2} and~\ref{soundfrequencyscaling2}, yields:
\begin{equation}\label{beta1}
\beta_{p,{\rm long}}(T)=\Big(\frac{L_S(T)}{L_S(T_0)}\Big)^{\frac{1}{2}}\Big(\frac{V(T)}{V(T_0)}\Big)^{\frac{1}{6}},
\end{equation}
\begin{equation}\label{beta2}
\beta_{p,{\rm tran}}(T)=\Big(\frac{G_S(T)}{G_S(T_0)}\Big)^{\frac{1}{2}}\Big(\frac{V(T)}{V(T_0)}\Big)^{\frac{1}{6}}.
\end{equation}
These equations describe the softening of the longitudinal and transverse acoustic components of the phonon density of states with temperature, based on elastic moduli.

Since the experimental phonon density of states often lacks separation into these components, we use a weighted average of Equations~\ref{beta1} and ~\ref{beta2} which accounts for the number of longitudinal acoustic and transverse acoustic phonon modes:
\begin{equation}\label{meanmodulus}
\beta_p^{-1}(T)\approx\Biggl(\frac{1}{3}\bigg(\frac{L_S(T_0)}{L_S(T)}\bigg)^{\frac{1}{2}}+\frac{2}{3}\bigg(\frac{G_S(T_0)}{G_S(T)}\bigg)^{\frac{1}{2}} \Biggr)^\eta \Big(\frac{V(T)}{V(T_0)}\Big)^{-\frac{1}{6}}.
\end{equation}
Note, this expression will be dominated by contributions from $G_S(T)$. Our reason for taking the mean of the reciprocal $\beta_p^{-1}(T)$, as opposed to $\beta_p(T)$, is that it is the reciprocal quantity that scales the phonon density of states in Equation~\ref{phononDOS}.

The changes in phonon frequency with temperature described by Equations~\ref{beta1} -~\ref{meanmodulus} strictly apply in the limit $\omega\rightarrow0$ because they are derived from changes in sound velocities. However, phonon modes other than those probed by sound velocities (\textit{e.g.}, high-frequency phonon modes) may soften at different rates with temperature. To capture these contributions to the phonon entropy, we have introduced the exponent $\eta$ in Equation~\ref{meanmodulus} as a corrective factor. This exponent allows for variance in softening between high and low-frequency phonons, while maintaining $\beta_p(T_0)=1$ at $T_0$. The choice of $\eta$ reflects the frequency distribution of phonon softening, and is found (below) to be correlated with the Poisson's ratio.

\begin{figure}
\begin{center}
\includegraphics[width=0.95\linewidth]{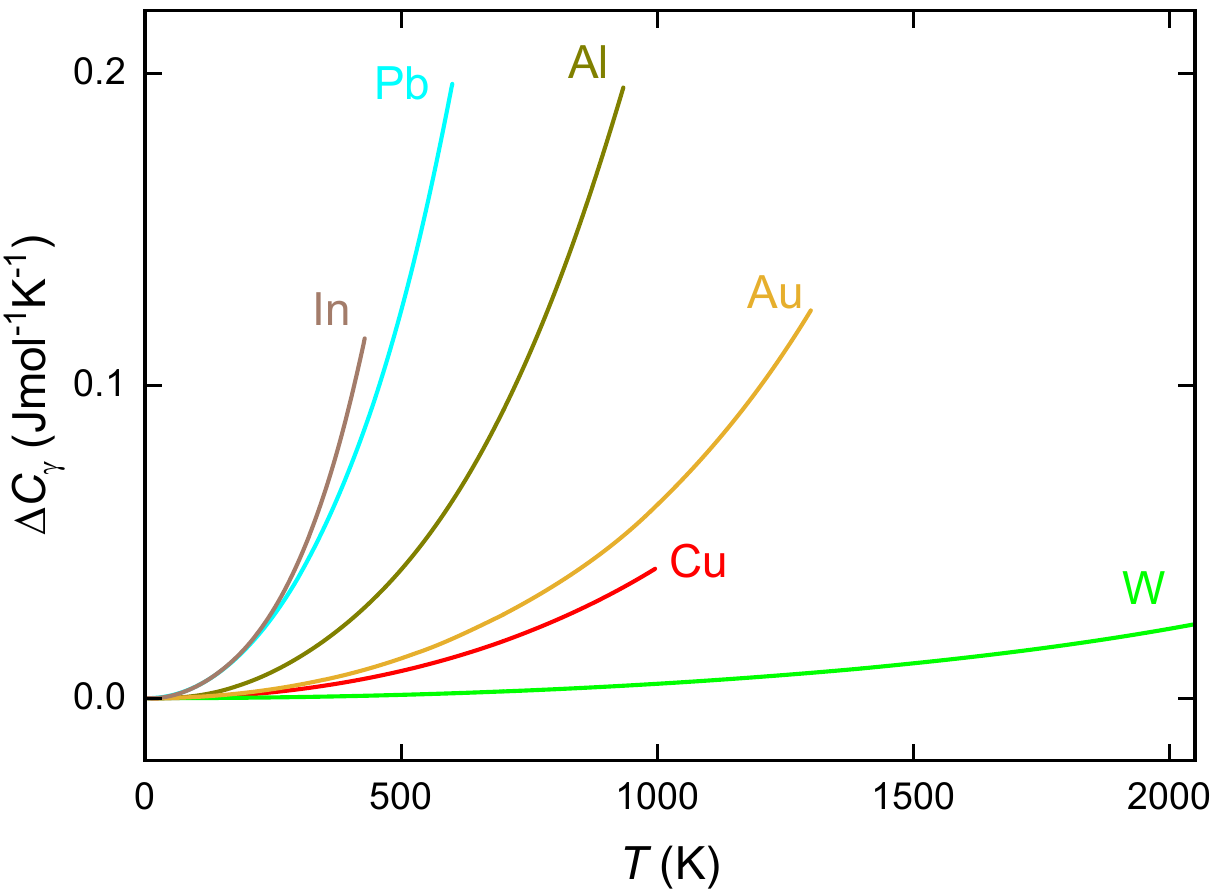}
\vspace{-4mm}
\textsf{\caption{
\textcolor{black}{Electronic Gr\"{u}neisen contribution to the specific heat ($\Delta C_\gamma=\gamma_{\mathrm{el}}\,\Gamma_{\mathrm{el}}\,\alpha_V(T)\,T^2$) used in the ESA. Values of $\gamma_{\rm el}$ are listed in Table~\ref{table2}.}
}
\vspace{-4mm}
\label{gruneisen}}
\end{center}
\end{figure}

\begin{figure*}
\begin{center}
\includegraphics[width=0.8\linewidth]{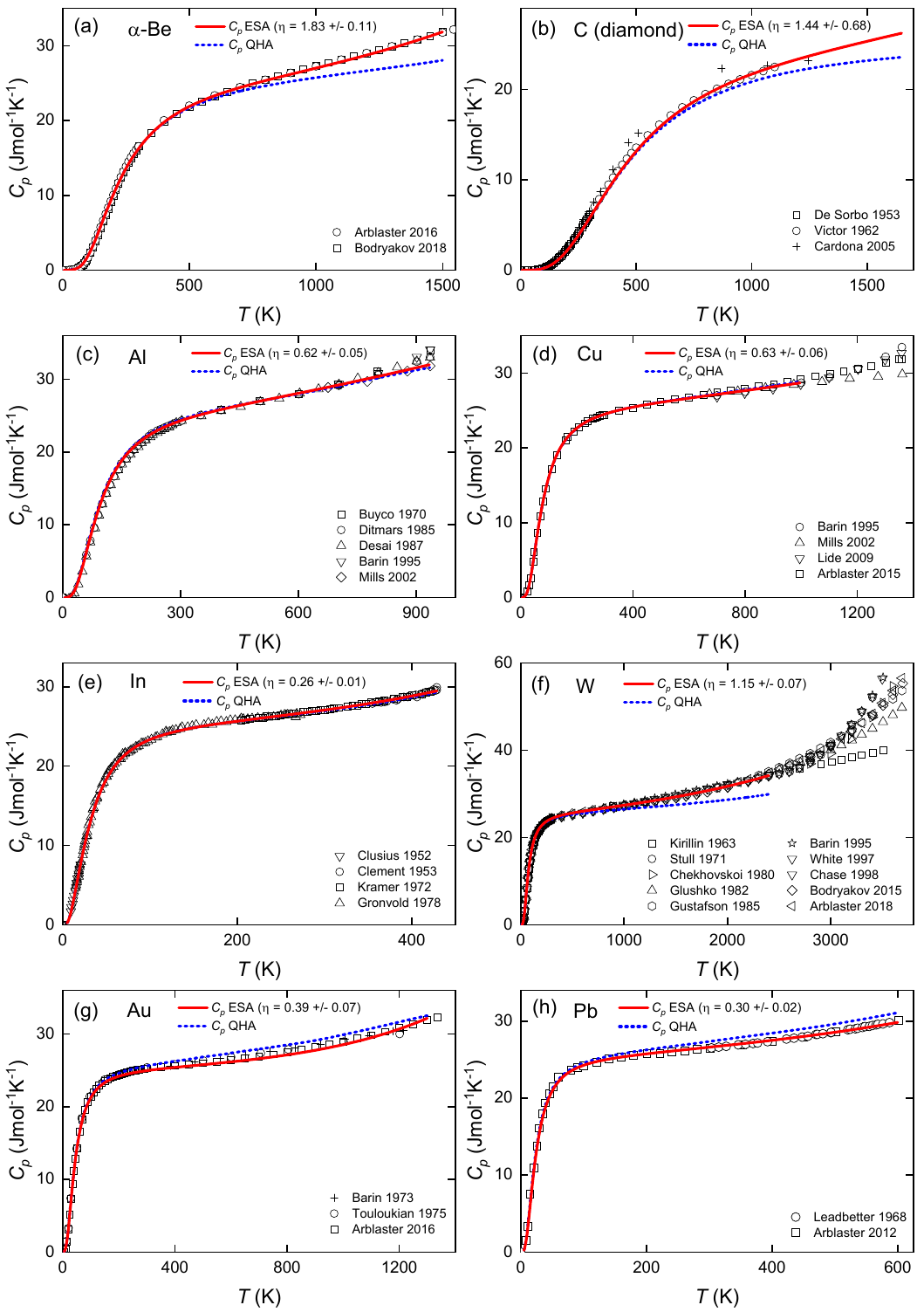}
\vspace{-4mm}
\textsf{\caption{
Specific heat data (represented by symbols) for $\alpha$-Be (a), C (diamond) (b), Al (c), Cu (d), In (e), W (f), Au (g), and Pb (h) from various sources~\cite{arblaster2016,bodryakov2018,desorbo1953, victor1962, cardona2005, buyco1970, ditmars1985, desai1987, barin1995, mills2002, chase1998, lide2010, arblaster2015, clusius1952, clement1953, kramer1972, gronvold1978, kirillin1963, stull1971, glushko1982, white1997, bodryakov2015, arblaster2018, chekhovskoi1980, gustafson1985,arblaster2016i, touloukian1975, barin1973,leadbetter1968,arblaster2012}, as indicated. The dotted black lines represent $C_p^{\rm QHA}(T)$ calculated assuming the quasiharmonic approximation (QHA) using Equation~\ref{quasiharmonic2}. The solid red lines represent $C_p^{\rm ESA}(T)$ calculated assuming the elastic softening approximation (ESA) using Equation~\ref{elasticsoftentingapproximation2}.
}
\vspace{-4mm}
\label{specificheat}}
\end{center}
\end{figure*}

\subsection{Accounting for Electronic Entropy}

\textcolor{black}{Since the total heat capacity of a metallic elemental solid includes an electronic part, it must also be quantified. For conventional metals, one typically adopts the Sommerfeld approximation, $C_{V,\mathrm{el}}(T) = \Gamma_{\mathrm{el}}\,T$, where $\Gamma_{\mathrm{el}}$ is the Sommerfeld coefficient~\cite{ashcroft1976}. Under the harmonic approximation, the complete specific heat becomes
\begin{equation}\label{elecharmonic2}
C_V^{\mathrm{HA}}(T) \;=\; C_{V,\mathrm{ph}}^{\mathrm{HA}}(T) + \Gamma_{\mathrm{el}}\,T.
\end{equation}
As with the phonon contribution, the volume dependence of the electronic term is not taken into account.}

\textcolor{black}{In the case of the quasiharmonic approximation, it is instructive to consider the volume derivative of the entropy at constant temperature within Equation~\ref{quasiharmonic0}. Following Varley~\cite{varley1956}, this entropy derivative can be separated into phononic and electronic components:
\begin{equation}\label{varley}
\alpha_V(T)B_T(T)\equiv
\frac{\partial S}{\partial V}\bigg|_T=\frac{\partial S_{\rm ph}}{\partial V}\bigg|_T+\frac{\partial S_{\rm el}}{\partial V}\bigg|_T.
\end{equation}
This means that the volume-dependence of the electronic entropy is implicitly included in the measured $\alpha_V(T)$ and $B_T(T)$ in the quasiharmonic approximation (see Appendix A). Hence, for the quasiharmonic approximation, the complete specific heat becomes   
\begin{equation}\label{quasiharmonic2}
C_p^{\mathrm{QHA}}(T) = C_{V,\mathrm{ph}}^{\mathrm{HA}}(T) +V(T)\alpha_V^2(T)B_T(T)T+ \Gamma_{\mathrm{el}}T.
\end{equation}
The simplicity of implementing Equation~\ref{quasiharmonic2} is the key reason why the quasiharmonic approximation is widely implemented.} 

\textcolor{black}{In the elastic softening approximation, the phonon contribution to the specific heat is considered to be
\begin{equation}
C_{p,\mathrm{ph}}^{\mathrm{ESA}}(T)
\;=\;
T\,\frac{\partial S_{\rm ph}}{\partial T}\bigg|_V
\;+\;
V(T)\,\alpha_V(T)\,T\,\frac{\partial S_{\rm ph}}{\partial V}\bigg|_T,
\end{equation}
which comprises two parts. The first term on the right-hand side involves the derivative of the phonon entropy with respect to temperature and therefore includes changes in thermal occupancy and phonon-frequency softening caused by intrinsic anharmonicity effects. The second term includes the same volume-derivative component of the phonon entropy that appears in the quasiharmonic approximation (and is contained on the right-hand side of Equation~\ref{varley}).} 


\textcolor{black}{Because the elastic softening approximation, unlike the quasiharmonic approximation, does not implicitly include the volume dependence of the electronic entropy ({\it i.e.}, the second term in Equation~\ref{varley}), it must be added as a separate term.  To approximate this electronic contribution, one can introduce an {electronic} Gr\"uneisen parameter $\gamma_{\mathrm{el}}$, defined by~\cite{varley1956}
\begin{equation}\label{electronicgruneisen}
\frac{\partial S_{\mathrm{el}}}{\partial V}\bigg|_T
\;=\;
\frac{\gamma_{\mathrm{el}}\,\Gamma_{\mathrm{el}}\,T}{V(T)},
\end{equation}
where $\Gamma_{\mathrm{el}}$ is, once again, the standard Sommerfeld coefficient. Incorporating this approximation yields the complete expression for the ESA constant-pressure specific heat:
\begin{equation}\label{elasticsoftentingapproximation2}
C_p^{\mathrm{ESA}}(T)
\;=\;
C_{p,\mathrm{ph}}^{\mathrm{ESA}}(T)
\;+\;
\Gamma_{\mathrm{el}}\,T
\;+\;
\gamma_{\mathrm{el}}\,\Gamma_{\mathrm{el}}\,\alpha_V(T)\,T^2. 
\end{equation}
Although $\gamma_{\mathrm{el}}$ has been measured in only a few materials (see Table~\ref{table2})---and typically at very low temperatures~\cite{barron1980}---its overall effect on $C_p(T)$ appears to be minor. Even at the highest temperatures, the resulting contribution to $C_p(T)$, $\Delta C_\gamma=\gamma_{\mathrm{el}}\,\Gamma_{\mathrm{el}}\,\alpha_V(T)\,T^2$, remains below approximately $0.2\:\mathrm{J\,mol^{-1}\,K^{-1}}$ (see Fig.~\ref{gruneisen}).}

\textcolor{black}{Other possible contributions to the electronic specific heat include the variation of $\Gamma_{\mathrm{el}}$ with temperature as $T$ approaches the Fermi temperature~\cite{ashcroft1976}, as well as potential electronic anharmonicity effects. If these effects exist, they must be incorporated into all specific-heat approximations---the harmonic approximation, quasiharmonic approximation, and elastic softening approximation---and thus would not affect estimates of the impact of {phonon} anharmonicity on the specific heat.}

%

\section{Application to Elemental Solids}

\subsection{Regular Elemental Solids}

Figure~\ref{specificheat} compares the calculated values of \(C_p^{\rm QHA}(T)\) and \(C_p^{\rm ESA}(T)\) using the quasiharmonic and elastic softening approximations against experimental specific heat data for various non-actinide elemental solids which adopt a range of crystal \textcolor{black}{and electronic} structures: \(\alpha\)-Be~\cite{arblaster2016, bodryakov2018}, diamond~\cite{desorbo1953, victor1962, cardona2005}, Al~\cite{buyco1970, ditmars1985, desai1987, barin1995, mills2002}, Cu~\cite{barin1995, chase1998, mills2002, lide2010, arblaster2015}, In~\cite{clusius1952, clement1953, kramer1972, gronvold1978}, W~\cite{kirillin1963, stull1971, glushko1982, white1997, chase1998, bodryakov2015, arblaster2018, chekhovskoi1980, gustafson1985, barin1995}, Au~\cite{arblaster2016i, touloukian1975, barin1973}, and Pb~\cite{leadbetter1968,arblaster2012}.

For the quasiharmonic approximation, the experimental volume, thermal expansivity, and isothermal bulk modulus data used to calculate \(C_p^{\rm QHA}(T)\) via Equation~\ref{quasiharmonic2} are detailed in Appendix A. The phonon density of states measurement temperatures (\(T_0\)), along with the volume and isothermal bulk moduli at these temperatures, are listed in Table~\ref{table2}. The Sommerfeld coefficients~\cite{stewart1983} employed in Equation~\ref{quasiharmonic2} are also provided in Table~\ref{table2}.

Figure~\ref{specificheat} demonstrates that the quasiharmonic approximation aligns well with the experimental behavior of \(C_p(T)\) for elements like Al, Cu, and In. However, for most solids, particularly at higher temperatures, this approximation fails to match the experimental curves. For instance, in cases like \(\alpha\)-Be, diamond, and W, the quasiharmonic approximation significantly underestimates \(C_p(T)\), whereas, for Au and Pb, it overestimates \(C_p(T)\).

\begin{figure}
\begin{center}
\includegraphics[width=0.95\linewidth]{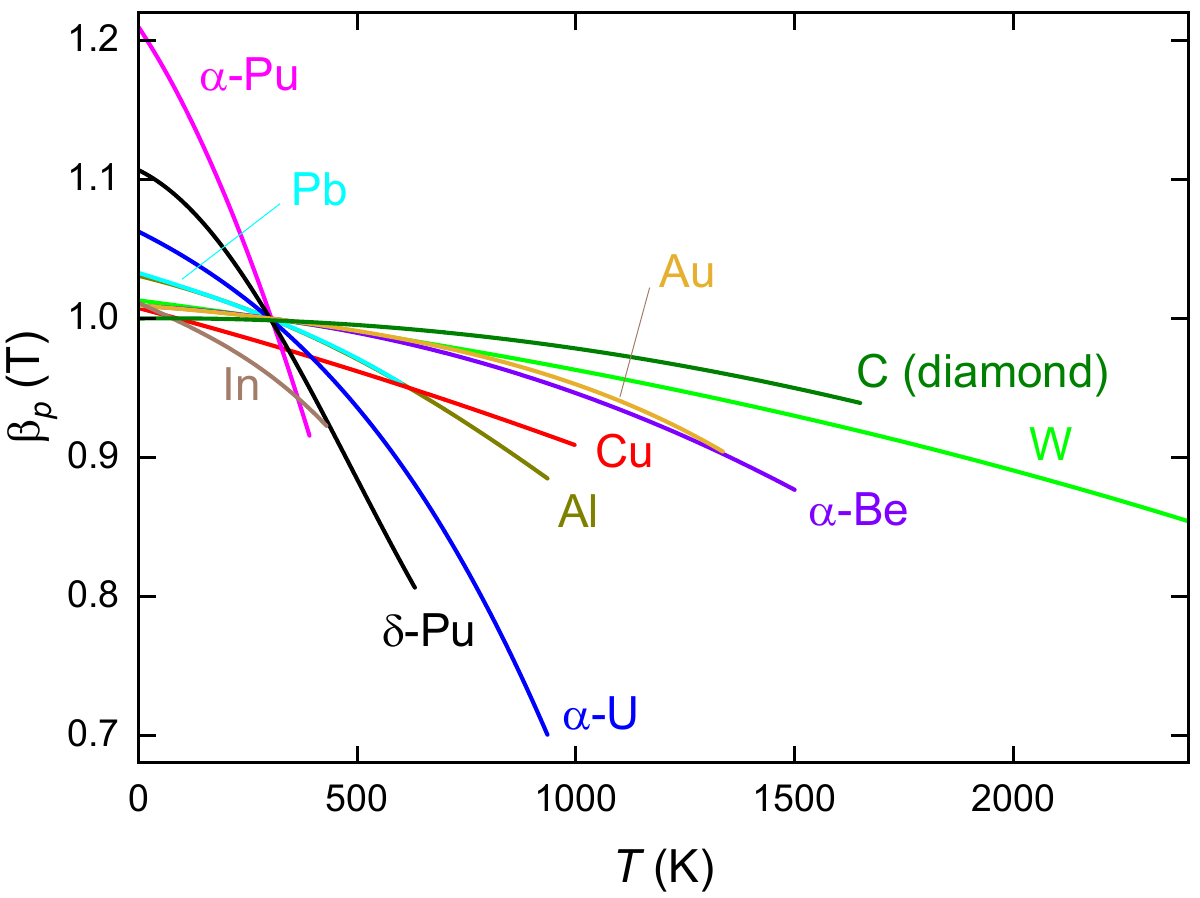}
\vspace{-4mm}
\textsf{\caption{
The softening parameter $\beta_p(T)$ for each elemental solid determined using Equation~\ref{meanmodulus}. Each line corresponds to a different elemental solid, as indicated.
}
\vspace{-4mm}
\label{betasofteningfig}}
\end{center}
\end{figure}

Conversely, our elastic softening approximation has successfully replicated the observed behavior of \(C_p(T)\) for all elemental solids in Fig.~\ref{specificheat} by finding specific values of \(\eta\) (listed in Table~\ref{table2}). The calculated \(\beta_p(T)\) curves, derived from Equation~\ref{meanmodulus}, are displayed in Fig.~\ref{betasofteningfig}. These values of \(\eta\), along with their margins of error determined via a least squares analysis described in Appendix B, suggest that the elastic softening approximation provides a more accurate model for specific heat behavior, particularly at high temperatures. The adiabatic bulk and shear moduli, \(B_S(T)\) and \(G_S(T)\), respectively, used for each elemental solid, are shown in Fig.~\ref{shear}, with the rationale for their selection detailed in Appendix C. \textcolor{black}{Meanwhile, values of $\gamma_{\rm el}$ used to determine the electronic Gr\"{u}neisen contribution in Equation~\ref{elasticsoftentingapproximation2} are listed in Table~\ref{table2}}.

\textcolor{black}{It is especially noteworthy that the elastic softening approximation performs well for W, which is known to exhibit a large anharmonic contribution to the entropy near its melting temperature~\cite{wallace1997} ($T_{\rm m}\approx$~3695~K). However, our comparison between the quasiharmonic and elastic softening approximations is limited by existing elastic moduli measurements for W, which only extend to 2400~K. We speculate that the upturn in the specific heat of W is related primarily to the softening of the shear modulus. 
}

\begin{figure*}
\begin{center}
\includegraphics[width=0.65\linewidth]{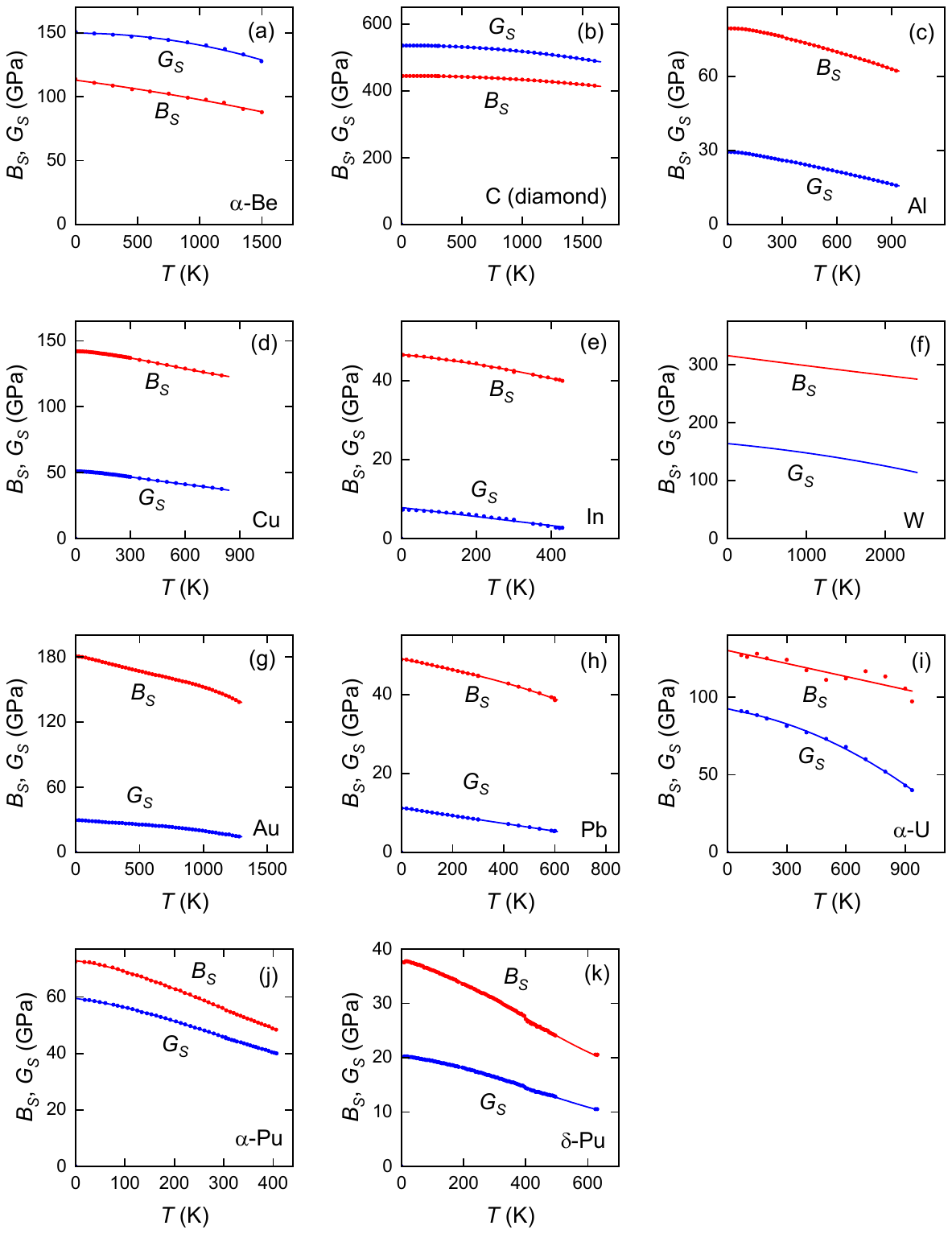}
\vspace{-4mm}
\textsf{\caption{
Adiabatic bulk modulus $B_S(T)$ and shear modulus $G_S(T)$ for $\alpha$-Be~\cite{dremov2009} (a), C (diamond)~\cite{zouboulis1998,migliori2008} (b), Al\cite{kamm1964,gerlich1969} (c), Cu~\cite{overton1955,chang1966} (d), In~\cite{chandrasekhar1961,vold1977} (e), W~\cite{lowrie1967} (f), Au~\cite{neighbours1958,chang1966} (g), Pb~\cite{waldorf1962,vold1977} (h), $\alpha$-U~\cite{armstrong1972} (i), $\alpha$-Pu~\cite{suzuki2011} (j), and $\delta$-Pu~\cite{suzuki2011,freibert2012} (k) from various sources, as indicated. For $\delta$-Pu, measurements below 500~K pertain to samples where the $\delta$ phase has been stabilized down to low temperatures by adding 2 atomic percent Ga. Circles represent data points (see Appendix C for details). To bridge gaps between data points and minimize the impact of scatter in the experimental data on thermodynamic derivatives, $B_S(T)$ and $G_S(T)$ are approximated with polynomial fits (solid lines).
}
\vspace{-4mm}
\label{shear}}
\end{center}
\end{figure*}

\subsection{Actinide Elemental Solids}

\begin{figure}
\begin{center}
\includegraphics[width=0.95\linewidth]{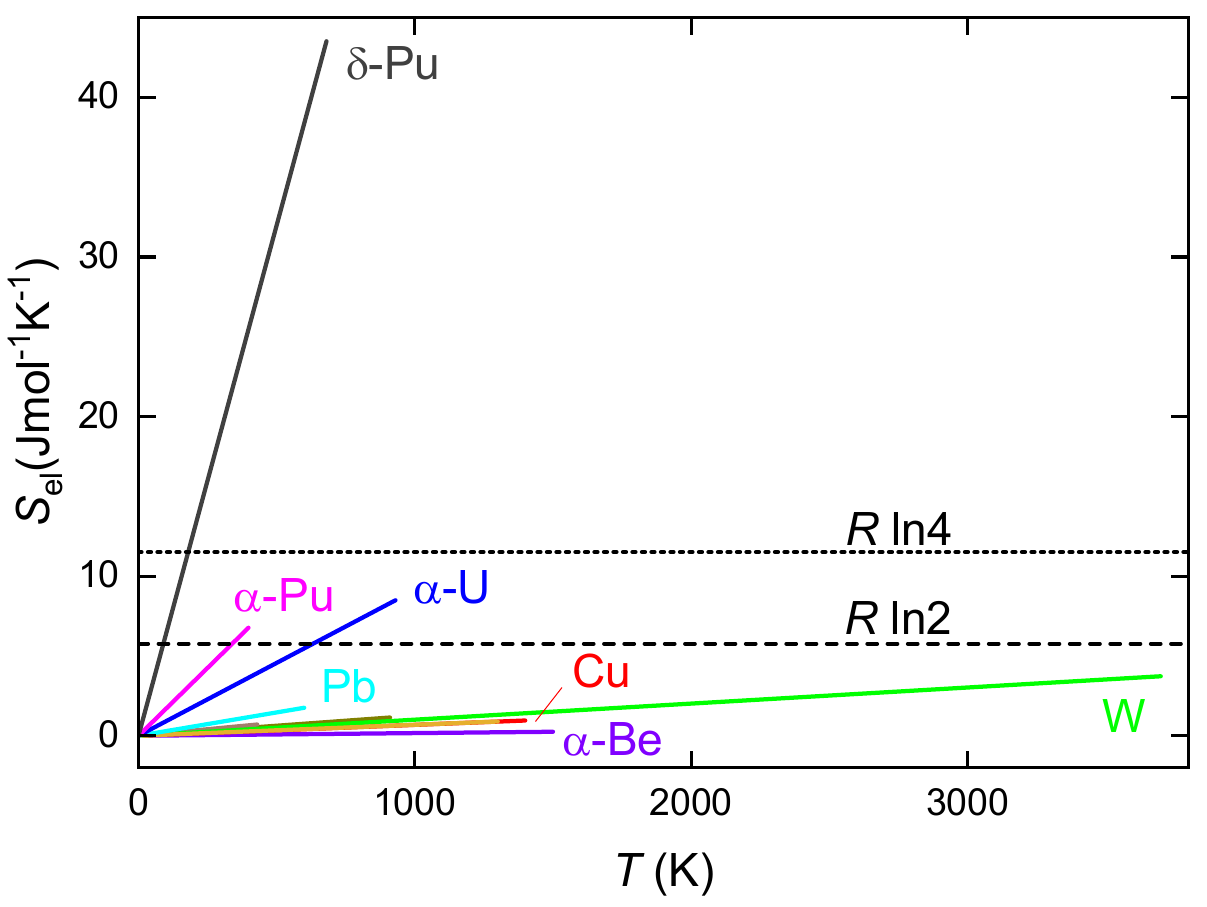}
\vspace{-4mm}
\textsf{\caption{
Estimated Sommerfeld contribution to the entropy $S_{\rm el}(T)$ under the assumption that the electronic contribution to $C_p(T)$, as given by Equations~\ref{quasiharmonic2} and~\ref{elasticsoftentingapproximation2}, continues to follow the Sommerfeld form at elevated temperatures. \textcolor{black}{For visual clarity, some element labels have been omitted}.
}
\vspace{-4mm}
\label{sommerfeldfig}}
\end{center}
\end{figure}

An important characteristic distinguishing actinide elemental solids from non-actinides, as observed in Table~\ref{table2}, is their notably larger Sommerfeld coefficients~\cite{lashley2003,lashley2001}. In actinides, if the electronic entropy were to follow the conventional form $S_{\rm el}\approx\Gamma_{\rm el}T$, Fig.~\ref{sommerfeldfig} suggests that at higher temperatures, the entropy would reach or exceed the maximum values typical for an ideal half-filled electronic band ($R\ln4$) or for a lattice of twofold degenerate spins ($R\ln2$), characteristic of a Kondo lattice system~\cite{hewson1993}. This observation implies that the assumption of a constant $\Gamma_{\rm el}$ becomes untenable in actinides like $\alpha$-U, $\alpha$-Pu, and $\delta$-Pu at elevated temperatures.

\begin{figure}
\begin{center}
\includegraphics[width=0.85\linewidth]{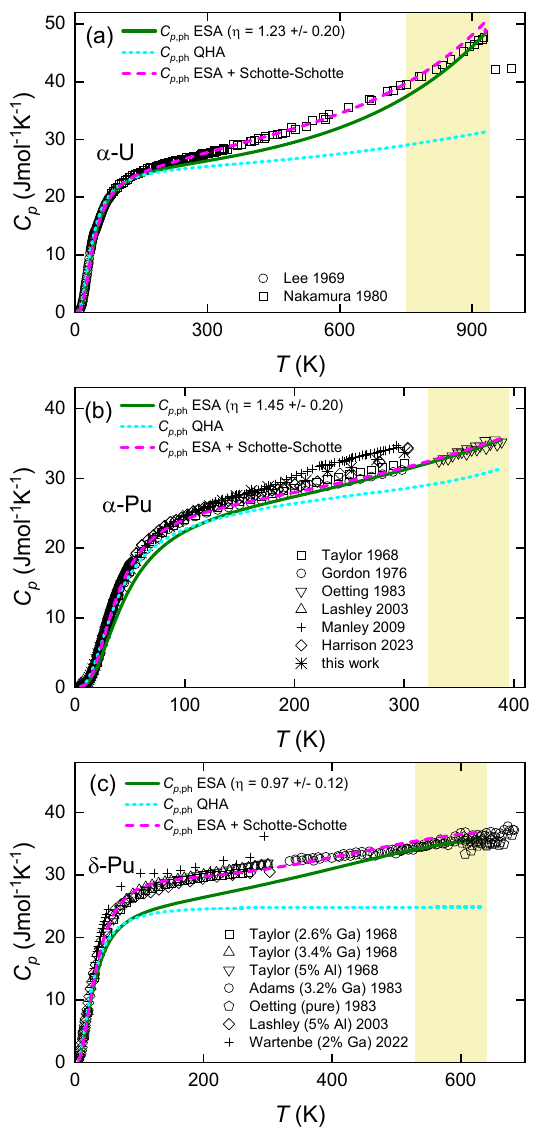} 
\vspace{-4mm}
\textsf{\caption{
Specific heat data (represented by symbols) for $\alpha$-U (a), $\alpha$-Pu (b), and $\delta$-Pu (c), sourced from various references~\cite{lee1969,nakamura1980, taylor1968, gordon1976, oetting1983, lashley2003, manley2009, harrison2023, adams1983, wartenbe2022}, as indicated. The dotted cyan lines represent $C_{p,{\rm ph}}^{\rm QHA}(T)$ calculated under the quasiharmonic approximation (QHA) using Equation~\ref{quasiharmonic}. The solid green lines represent $C_{p,{\rm ph}}^{\rm ESA}(T)$ calculated under the elastic softening approximation (ESA) using Equation~\ref{phononenytropySA}. The dashed magenta curves illustrate $C_{p,{\rm ph}}^{\rm ESA}(T)$ combined with a contribution from a Schotte-Schotte anomaly, demonstrating the potential contribution from electronic or magnetic degrees of freedom. The yellow shaded region indicates high temperatures, which are the focus of our analysis (see text).
}
\vspace{-4mm}
\label{actinidespecificheat}}
\end{center}
\end{figure}

For $\delta$-Pu, various studies have suggested contributions to $C_p(T)$ from sources beyond phonons, such as local moments or narrow electronic bands~\cite{wartenbe2022,harrison2023,lawson2019,lawson2013}. These contributions significantly impact the entropy, with their effects prominent below room temperature and diminishing at higher temperatures. Similar observations have been made for $\alpha$-Pu~\cite{harrison2023}. In the case of $\alpha$-U, the $\Gamma_{\rm el}$ value in Table~\ref{table2} is linked to a phase with a reconstructed Fermi surface due to charge-density waves below approximately 40 K~\cite{lashley2001}, suggesting a significant residual $\Gamma_{\rm el}$ value in the absence of this order. Despite the nature—magnetic or electronic—of these additional entropy sources in the actinides, the high $\Gamma_{\rm el}$ values suggest an electronic entropy saturation at lower temperatures, thereby excluding a substantial electronic contribution to $C_p(T)$ at elevated temperatures \textcolor{black}{(the region shaded in yellow in Fig.~\ref{actinidespecificheat})}.


In Fig.~\ref{actinidespecificheat}, we compare $C_{p,{\rm ph}}^{\rm ESA}(T)$ and $C_{p,{\rm ph}}^{\rm QHA}(T)$ against experimental data for $\alpha$-U~\cite{lee1969,nakamura1980}, $\alpha$-Pu~\cite{taylor1968, gordon1976, oetting1983, lashley2003, manley2009, harrison2023}, and $\delta$-Pu~\cite{oetting1983}. In pure Pu, the $\delta$ phase exists only within a temperature range of approximately \textcolor{black}{588~K to 723~K}~\cite{jette1954, schonfeld1996}. We have therefore included data from Pu samples in which the $\delta$ phase is stabilized down to lower temperatures by the substitution of 2 or more atomic percent of the Pu atomic sites with Al or Ga~\cite{hecker2004}, as indicated~\cite{adams1983, taylor1968, lashley2003, harrison2023, wartenbe2022}. Our analysis is consistent with a minor electronic contribution at elevated temperatures, which we illustrate by approximating contributions from electronic or magnetic degrees of freedom in actinides with a Schotte-Schotte anomaly~\cite{schotte1975} (magenta curves in Fig.~\ref{actinidespecificheat}), similar to those discussed in Ref.~\cite{harrison2023} (see Appendices D and E). The electronic contribution to $C_p(T)$ in the yellow shaded region at high temperatures (from the \textcolor{black}{Schotte-Schotte} anomaly at low temperatures) is shown to be small.

By utilizing $\eta$ values consistent with those of regular elemental solids (Table~\ref{table2}), the elastic softening approximation accurately reflects the experimentally observed specific heat, unlike the quasiharmonic approximation, which significantly underestimates the phonon contribution at high temperatures and overestimates it at low temperatures. The differences in the performance of these two approximations are further elucidated in Appendix D through a residuals analysis.

\subsection{Validation of the Elastic Softening Approximation}

\begin{figure}
\begin{center}
\includegraphics[width=0.85\linewidth]{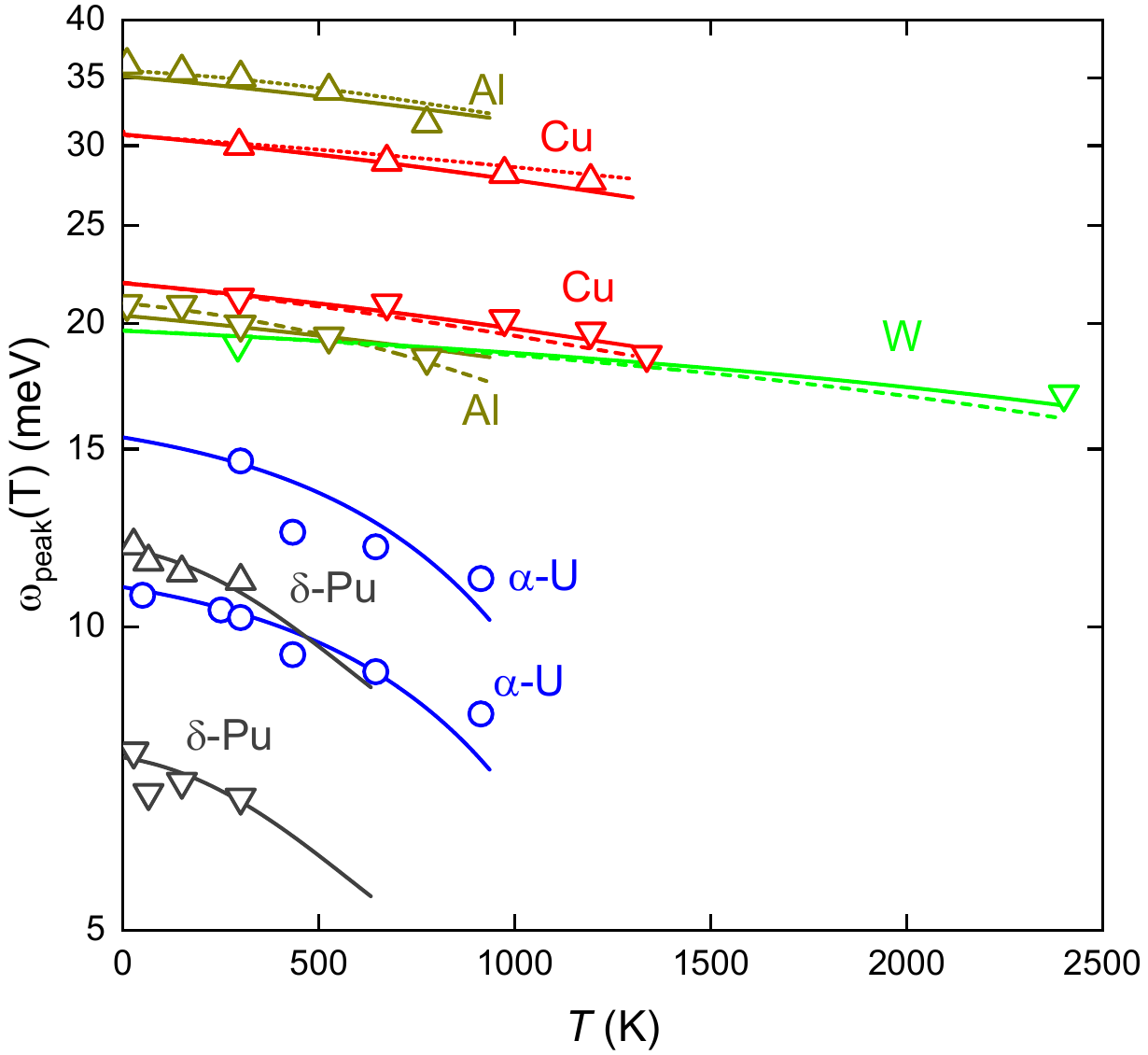}
\vspace{-4mm}
\textsf{\caption{
The measured temperature dependences of prominent features in the phonon density of states (DOS) determined by neutron scattering~\cite{kresch2008,larose1976, semenov2014,manley2001,mcqueeney2004} for Al, Cu, W, $\alpha$-U, and $\delta$-Pu, as indicated by symbols. Solid lines represent $\omega_{\rm peak}(T)=\omega_{\rm peak}(T_0)\beta_p(T)$, where $\omega_{\rm peak}(T_0)$ is used to rescale $\beta_p(T)$ from Fig.~\ref{betasofteningfig}. For Cu, a spline interpolation is used between 997~K and 1300~K. Circles are used to identify features in the phonon DOS that are attributed primarily to optical phonon modes~\cite{crummett1979}. Upward-pointing triangles denote features in the DOS predominantly associated with longitudinal phonon modes, while dotted lines indicate $\omega_{\rm peak}(T)=\omega_{\rm peak}(T_0)\beta_{p,{\rm long}}^\eta(T)$. Downward-pointing triangles indicate features primarily arising from transverse phonon modes, while dashed lines indicate $\omega_{\rm peak}(T)=\omega_{\rm peak}(T_0)\beta_{p,{\rm tran}}^\eta(T)$. 
}
\vspace{-4mm}
\label{frequencyshiftfig}}
\end{center}
\end{figure}

Beyond reproducing the temperature dependence of the specific heat in a range of elements solids, there are two key observations that support the accuracy of our elastic softening approximation in understanding anomalous phonon contributions to the specific heat of elemental solids, particularly at elevated temperatures. First, we find consistency between the phonon frequency softening parameter $\beta_p(T)$ and the experimental temperature-dependent phonon density of states. Figure~\ref{frequencyshiftfig} shows the locations of maxima within the phonon density of states for various materials, obtained from neutron scattering studies~\cite{kresch2008,larose1976, semenov2014,manley2001,mcqueeney2004,crummett1979}.  \textcolor{black}{Note that in the case of Cu, the frequency shifts pertain to the maxima of the longitudinal and transverse phonon dispersions~\cite{larose1976}, which contribute maximally to the peaks in the phonon density of states. The locations of the maxima in the density of states conform well to $\omega_{\rm peak}(T)=\omega_{\rm peak}(T_0)\beta_p(T)$, where $\omega_{\rm peak}(T_0)$ is a constant used to rescale $\beta_p(T)$ from Fig.~\ref{betasofteningfig}. 
The ability of the phonon softening parameter $\beta_p(T)$ to successfully reproduce the temperature dependence of high-frequency portions of the phonon density of states lends credence to our approach of scaling low-frequency phonon softening, provided by elastic moduli measurements, to model the entire temperature-dependent phonon density of states.}

\textcolor{black}{In the cases of Al, Cu and W, for which maxima in the phonon density of states are known to originate from longitudinal or transverse branches, we also show dashed lines respectively  corresponding to $\omega_{\rm peak}(T)=\omega_{\rm peak}(T_0)\beta_{p,{\rm long}}^\eta(T)$ and $\omega_{\rm peak}(T)=\omega_{\rm peak}(T_0)\beta_{p,{\rm tran}}^\eta(T)$ (note the need to apply the exponent $\eta$ for equivalent scaling). The difference in softening behavior between the weighted average, $\omega_{\rm peak}(T)=\omega_{\rm peak}(T_0)\beta_p(T)$, and those of the longitudinal or transverse phonon branches is small.}

\textcolor{black}{In the case of $\alpha$-uranium, most phonon modes (both optical and acoustic) soften in the usual manner as the temperature increases~\cite{bouchet2015,bouchet2017}. However, there is an {anomalous} mode, referred to as $\Sigma_4$, that softens {as the temperature decreases}. This softening precedes the eventual condensation of that mode at low temperature in a charge-density wave  phase~\cite{smith1980}. Despite this unusual behavior, the general scaling relation
\(
\omega_{\text{peak}}(T) \;=\; \omega_{\text{peak}}(T_0)\,\beta_p(T)
\)
still appears to hold when looking at the overall phonon spectrum. The key point here is that the $\Sigma_4$ mode corresponds to a very small volume in reciprocal space. Its contribution to the total phonon density of states is therefore too small to create a readily visible peak or distinct feature in the phonon density of states~\cite{manley2001} or in the specific heat at higher temperatures. On the other hand, the continued influence of an incipient charge density wave on high-temperature properties may be tied to the observed pronounced softening of the shear modulus with increasing temperature~\cite{bouchet2017}. In a manner similar to W, we find this softening to be the principal cause of the strong upturn in $C_p$ with increasing temperature.}

\begin{figure}
\begin{center}
\includegraphics[width=0.85\linewidth]{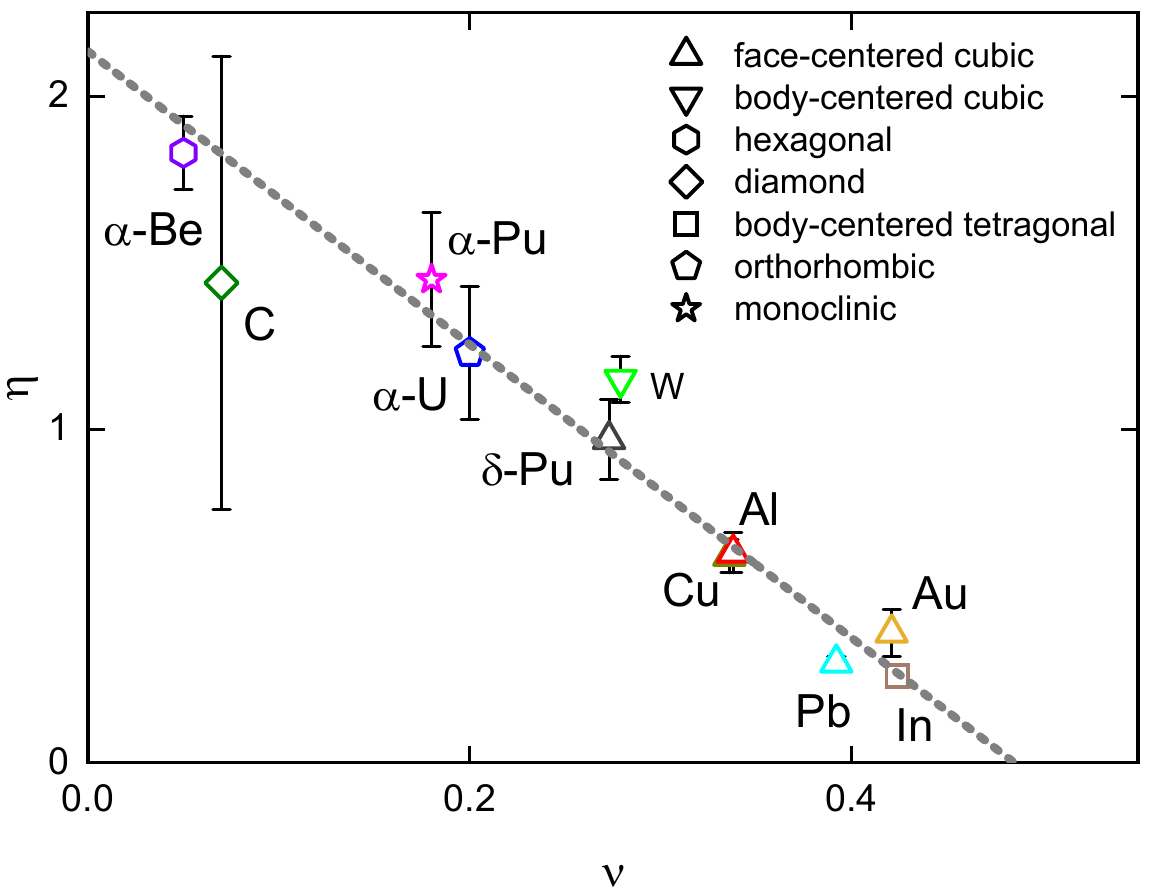}
\vspace{-4mm}
\textsf{\caption{
The differential softening parameter $\eta$, along with its error bars, is plotted against Poisson's ratio $\nu$ for each elemental solid. Different symbol shapes in the plot correspond to various crystal structures, as indicated. \textcolor{black}{The values of $\eta$ are also listed in Table~\ref{table2}}. 
}
\vspace{-4mm}
\label{poissonsratiofig}}
\end{center}
\end{figure}

\textcolor{black}{The second key observation supporting our elastic softening approximation approach is that} the values of $\eta$ for the elemental solids investigated within this work are not random, but correlated with the nature of the bonding in the elemental solids. This is evidenced by the correlation between $\eta$ and Poisson's ratio shown in Fig.~\ref{poissonsratiofig}. Poisson's ratio describes the transverse strain arising in response to a longitudinal strain~\cite{greaves2019}. In general this will be an anisotropic quantity, but for an isotropic crystal it is related to the bulk and shear elastic moduli through

\begin{equation}\label{poissonsratio}
\nu=\frac{3B_S-2G_S}{6B_S+2G_S}.
\end{equation}
Poisson's ratio is also temperature dependent, reflecting the temperature dependence of the underlying elasticity.

As a ratio of elastic moduli, $\nu$ is known to provide a good indication of the type of interatomic bonding present within a solid~\cite{koster1961,ledbetter2008}. For isotropic solids, three cases are typically considered~\cite{johnson1988, pettifor1992, eberhart2012}: (1) $\nu = 0.25$ corresponding to interatomic bonding mediated through a central force potential; (2) $\nu > 0.25$ indicating the presence of more metallic bonding character; (3) $\nu < 0.25$ pointing towards more covalent bonding character. Since it reflects bonding character, Poisson's ratio is also often used as a proxy for intrinsic ductility ($\nu > 0.25$) and brittleness ($\nu < 0.25$) ~\cite{pugh1954, pettifor1991, senkov2021}.

The Poisson's ratio values shown in Fig.~\ref{poissonsratiofig} were evaluated using Equation~\ref{poissonsratio} in the limit $T\rightarrow0$. The linear trend in Fig.~\ref{poissonsratiofig} reveals a physical relationship linking anomalous phonon contributions to $C_p(T)$ with the bonding character of solids. Notably, actinide solids also align with this trend, confirming our finding that phonon degrees of freedom largely account for their specific heat at elevated temperatures. The dotted line in Fig.~\ref{poissonsratiofig} indicates an approximately linear relationship:
\begin{equation}\label{etanu}
\eta(\nu)=b(\nu_0-\nu),
\end{equation}
with $\nu_0=$~0.485~$\pm$~0.034 and $b=$~4.40~$\pm$~0.30. The larger error bars for diamond reflect limited high-temperature data. This relationship between $\nu$ and $\eta$ suggests that metallic bonding character reduces the contribution of elastic moduli changes to the phonon softening parameter $\beta_p$ such that only the volumetric term remains in Equation~\ref{meanmodulus} when $\nu \approx 0.5$. Alternatively, the elastic moduli changes provide larger contributions to $\beta_p$ with increased covalent bonding character.

\subsection{Ascertaining the Anharmonic Contribution}

\begin{figure}
\begin{center}
\includegraphics[width=0.9\linewidth]{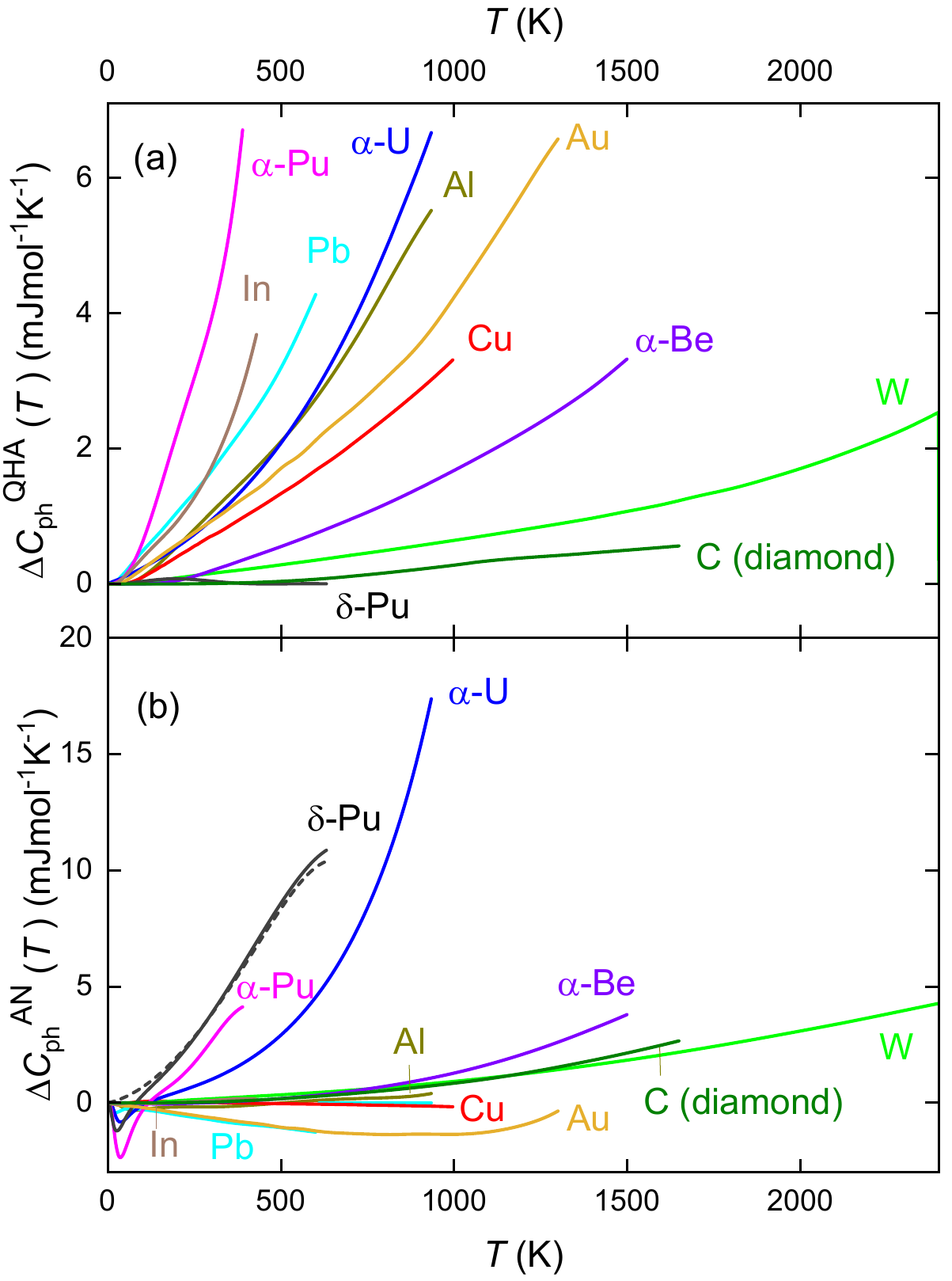}
\vspace{-4mm}
\textsf{\caption{
A comparison of $\Delta C^{\rm QHA}_{\rm ph}(T)$ (a) and $\Delta C^{\rm AN}_{\rm ph}(T)$ (b), as defined in the text. The dashed line in (b) represents $\Delta C^{\rm AN}_{\rm ph}(T)$ for $\delta$-Pu, estimated using $a(T)$ as given by Equation~\ref{anharmonicitydefinitiondelta}.
}
\vspace{-4mm}
\label{comparisonfig}}
\end{center}
\end{figure}

In Fig.~\ref{comparisonfig}, we present a comprehensive comparison of phonon contributions to the specific heat beyond the harmonic approximation for each of the elemental solids, including the actinides. Figure~\ref{comparisonfig}(a) illustrates the quasiharmonic contribution, denoted \textcolor{black}{$\Delta C^{\rm QHA}_{\rm ph}(T) = C_{p,{\rm ph}}^{\rm QHA}(T) - C_{V,{\rm ph}}^{\rm HA}(T) = V(T)\alpha_V^2(T)B_T(T)T-\gamma_{\mathrm{el}}\Gamma_{\mathrm{el}}\alpha_V(T)T^2$}. Figure~\ref{comparisonfig}(b) displays the additional anomalous softening contribution, \textcolor{black}{$\Delta C^{\rm AN}_{\rm ph}(T) = C_{p,{\rm ph}}^{\rm ESA}(T) - C_{p,{\rm ph}}^{\rm QHA}(T)+\gamma_{\mathrm{el}}\Gamma_{\mathrm{el}}\alpha_V(T)T^2$}, which is captured by our elastic softening approximation, but not by the quasiharmonic approximation.

\textcolor{black}{Recall that in the quasiharmonic approximation, the term $V(T)\,\alpha_V^2(T)\,B_T(T)\,T$ implicitly incorporates an electronic component. Although this electronic contribution is typically quite small (see Fig.~\ref{gruneisen}) and often omitted, we have corrected for this contribution by subtracting the corresponding Gr\"uneisen approximation, $\gamma_{\mathrm{el}}\Gamma_{\mathrm{el}}\alpha_V(T)\,T^2$, for conventional metals in Fig.~\ref{comparisonfig}. An analogous electronic Gr\"uneisen-like parameter (as in Equation~\ref{electronicgruneisen}) should, in principle, also exist in actinides. However, because the electronic contribution to their specific heat is strongly suppressed at high temperatures, any such Gr\"uneisen-like term would likewise be diminished. In $\delta$-Pu, this term is expected to be negligible, given its extremely small thermal expansivity.}

To the extent that $C^{\rm QHA}_{p,{\rm ph}}$ accurately captures the quasiharmonic contribution to the specific heat, it is reasonable to propose that $\Delta C^{\rm AN}_p(T)$ approximates the specific heat contribution due to intrinsic anharmonicity. This association is particularly robust in the case of $\delta$-Pu, where the volume change with temperature is minimal. This interpretation is supported by a standard definition of intrinsic anharmonicity~\cite{organov2004,dove2016}:
\begin{equation}\label{anharmonicitydefinition}
a(T)=\frac{\partial\ln\omega(T)}{\partial T}\bigg|_V.
\end{equation}
Given the near-constant volume of $\delta$-Pu, we can approximate that $\partial\omega(T)/\partial T|_V\approx\partial\omega(T)/\partial T|p$. This allows for the substitution $\beta_p(T)\propto{\omega(T)}$, as indicated in Equation~\ref{phononDOS}, leading to:
\begin{equation}\label{anharmonicitydefinitiondelta}
a(T)\approx\frac{\partial\ln\beta_p(T)}{\partial T}.
\end{equation}
Using the approximation $\Delta C_p^{\rm AN}=3a(T)RT$ for intrinsic anharmonic phonon contributions to the specific heat~\cite{wallace1972,organov2004}, we arrive at the dashed line in Fig.~\ref{comparisonfig}(b). This line is consistent with the form obtained using $\Delta C^{\rm AN}(T) = C_{p,{\rm ph}}^{\rm ESA}(T) - C_{p,{\rm ph}}^{\rm QHA}(T)$, thus validating our approach for quantifying the anharmonic contribution in $\delta$-Pu.

\section{Discussion}

A significant limitation of the quasiharmonic approximation is, despite its success in certain materials~\cite{fultz2010}, its neglect of intrinsic anharmonicity. Intrinsic anharmonicity is often considered to be important under extreme conditions of high temperature and pressure~\cite{jacobs2005}, but it can also be significant in scenarios where the thermal expansivity at constant pressure is small or negative~\cite{dove2016}. In the particular case of $\delta$-Pu, this oversight fails to capture the primary mechanism of phonon softening with temperature~\cite{mcqueeney2004}. Our proposed elastic softening approximation, based on an approach developed by Wallace~\cite{wallace1972}, effectively treats such scenarios.


A pivotal aspect of our approach is capturing the varying rates of softening of the phonon density of states at different frequencies, relative to the behavior observed in elastic moduli measurements which only directly probe the phonon density of states near $\omega\approx0$. To account for the thermodynamic effect of this differential softening at high temperatures, we have introduced the parameter $\eta$. This parameter is not arbitrary; it exhibits a pronounced correlation with Poisson's ratio (shown in Fig.~\ref{poissonsratiofig}). This finding suggests that $\eta$ exceeds one in materials with more covalent bonding character and is smaller in materials with more metallic bonding character. Given that Poisson's ratio is a well-established metric across all solids~\cite{koster1961,ledbetter2008,greaves2019}, we anticipate our elastic softening approximation can be used as a more accurate predictor of the phonon contribution to specific heat, inclusive of anharmonic effects, than the quasiharmonic approximation. Importantly, the correlation of $\eta$ with $\nu$ enables its use in predicting the specific heat profile over a broad temperature range without needing adjustable parameters.

The robustness of our method is further endorsed by the strong correlation between the overall phonon softening, $\beta_p(T)$, and the softening of high-energy spectral features in the phonon density of states, as inferred from neutron scattering measurements (shown in Fig.~\ref{frequencyshiftfig}). The parallel trends in the softening of both longitudinal and transverse modes at high frequencies justify our method's assumptions and the application of the weighted average approximation in Equation~\ref{meanmodulus}.

A critical aspect of our study is the use of the actual phonon density of states (as presented in Fig.~\ref{phononDOSfig}) and the measured temperature-dependent elastic moduli (Fig.~\ref{shear}), rather than approximations based on the Debye or Einstein models~\cite{ashcroft1976}. \textcolor{black}{Although Debye- or Einstein-based models can, for certain materials, help elucidate key physical properties~\cite{dimitrov2010,rathore2019}, this is no longer the case for systems displaying pronounced intrinsic anharmonicity. For instance, Debye- or Einstein-based approaches can fit the temperature dependence of $C_p(T)$ by using multiple adjustable parameters, yet they may obscure the actual extent of phonon softening and the novel underlying physics (as revealed in Figs.~\ref{frequencyshiftfig}, \ref{poissonsratiofig}, and~\ref{comparisonfig}).}

The findings presented in Fig.~\ref{comparisonfig} indicate that an intrinsic anharmonic phonon contribution to the specific heat is not only significant in $\delta$-Pu, but also prominently present in all three actinide solids we studied. It is crucial to emphasize that in these materials, electronic degrees of freedom primarily contribute to the specific heat at low temperatures (see Appendices D and E). This deviates from typical elemental solids where the Sommerfeld approximation usually suffices~\cite{ashcroft1976}.


\textcolor{black}{Because $\delta$-Pu has minimal thermal expansion, it allows a quantifiable analysis of its intrinsic anharmonic contribution to the specific heat, as illustrated in Fig.~\ref{comparisonfig}, in agreement with prior theoretical estimates~\cite{lawson2019,bottin2024}. For $\alpha$-Pu, our findings agree with earlier work indicating anharmonicity~\cite{filanovich2015,lawson2019}, while our results for $\alpha$-U likewise align with previous reports of anharmonicity~\cite{bouchet2017,bouchet2015} or anomalous phonon softening~\cite{manley2001} in that system.}

\textcolor{black}{We note that although one theoretical model predicts negligible anharmonic effects in $\delta$-Pu~\cite{soderlind2023}, it does not propose an alternative framework to explain the anomalously large phonon softening observed in $\delta$-Pu~\cite{mcqueeney2004}, nor clarify why such softening should not contribute to the specific heat~\cite{lashley2003} (see Appendix~H).}

\textcolor{black}{Although our ESA assumes that phonon frequencies depend on temperature, it does not require that this dependence arise only from a conventional anharmonic interatomic potential~\cite{varley1956,barron1965,wallace1972,allen2015}. For instance, the negative thermal expansion observed in $\delta$-Pu is thought to originate at least partly from electronic or magnetic mechanisms~\cite{rudin2022,harrison2020,lawson2006,migliori2016}, which may also underlie its anomalous phonon softening~\cite{harrison2020}. In $\alpha$-U, similar mechanisms to those invoked in $\delta$-Pu~\cite{manley2001} have been invoked to explain the substantial temperature-induced changes in phonon modes; additionally, strong electron-phonon coupling~\cite{riseborough2007} must be relevant, given the charge density wave ground state in this system~\cite{smith1980,bouchet2015,bouchet2017}. Importantly, despite the complex interplay between phonons and electronic degrees of freedom, the correlation between the parameter $\eta$ and Poisson's ratio (see Fig.~\ref{comparisonfig}) remains valid.}

Our comparative analysis of temperature-induced changes in the elastic moduli of $\delta$-Pu and other face-centered cubic solids reveals a unique trait in $\delta$-Pu. In contrast to the typical behavior of face-centered cubic materials (see Fig.~\ref{shear}), where a more rapid reduction in shear modulus relative to bulk modulus is seen, $\delta$-Pu maintains an almost constant shear-to-bulk modulus ratio with increasing temperature~\cite{suzuki2011}. This characteristic, differing from expectations for a material with a crystal structure similar to highly ductile elements, implies a greater degree of covalency in $\delta$-Pu. This observation is in line with the larger $\eta$ value found for $\delta$-Pu compared to other face-centered cubic elemental solids. The resistance to shear deformation in $\delta$-Pu, reminiscent of materials like $\alpha$-Be and diamond known for their rigid covalent bonding, suggests a mechanism potentially related to the coupling of $5f$-electrons to the crystal lattice. This hypothesis gains further support from the similar temperature dependence of shear and bulk moduli in $\alpha$-Pu, as shown in Fig.~\ref{shear}, which exhibits greater brittleness than $\delta$-Pu. 


\textcolor{black}{To establish the practicality of the ESA approach, we have focused on elemental solids with abundant experimental data. To broaden its applicability, one would need to find an approximate way to include elemental solids with more limited experimental information. This is especially relevant for phases stabilized exclusively at high temperatures. For instance, in high-temperature actinide phases (e.g., $\beta$-Pu~\cite{jette1954,schonfeld1996,suzuki2011} and $\gamma$-U~\cite{manley2001}), the magnitude and form of the electronic contribution are more uncertain, since characterizing them typically requires some low-temperature measurements to be made.}

\textcolor{black}{A further question is whether the elastic softening approximation approach can be extended to multi-element systems. In actinide compounds, for example, the mass differences between elements is especially large, and can lead to a substantial gap opening between acoustic and optical phonon modes. A key issue in such systems is whether the parameter $\eta$ can still capture the distinct behaviors of high-temperature optical modes (which would significantly affect the specific heat) versus the acoustic modes that dominate elastic moduli measurements, or whether further refinements are needed. A particularly relevant and useful test case for future study would be UO$_2$~\cite{sanati2011,pang2014,bryan2019,noguere2020}, as it has already been the subject of both elastic moduli and temperature-dependent phonon density-of-states measurements.}


%

\section*{APPENDIX A: PARAMETERS USED IN THE QUASIHARMONIC APPROXIMATION}

\begin{figure}
\begin{center}
\includegraphics[width=0.95\linewidth]{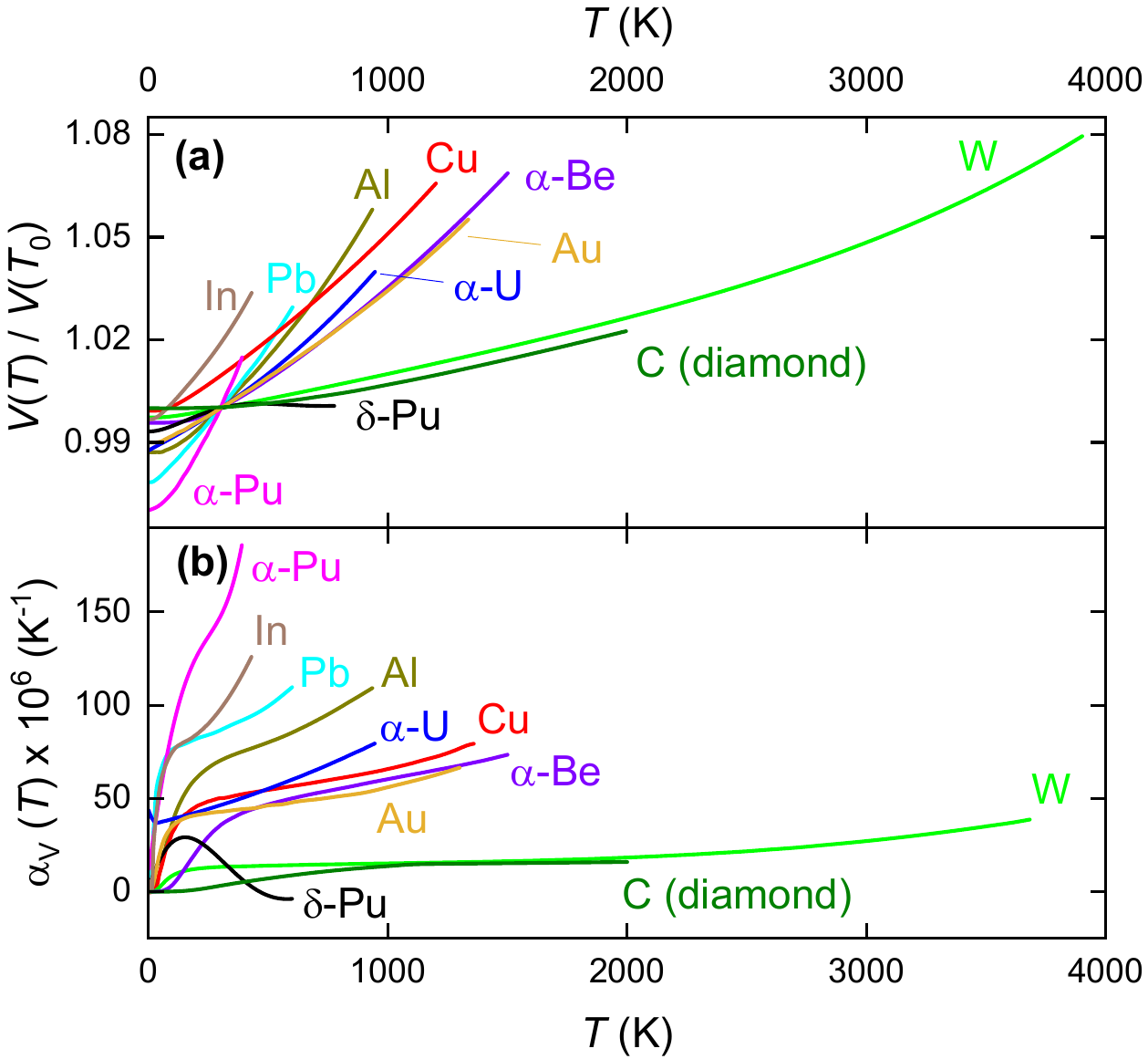}
\vspace{-4mm}
\textsf{\caption{Renormalized volume $V(T)/V(T_0)$ (a) and thermal expansivity $\alpha_V(T)$ (b) of $\alpha$-Be~\cite{bodryakov2018}, C (diamond)~\cite{jacobson2019}, Al~\cite{najashima2018,kozyrev2022}, Cu~\cite{white1997,kozyrev2023}, In~\cite{arblaster2018i}, W~\cite{kozyrev2023i}, Au~\cite{pamato2018}, Pb~\cite{kozyrev2022i}, $\alpha$-U~\cite{lloyd1966}, $\alpha$-Pu~\cite{harrison2023} and $\delta$-Pu~\cite{harrison2023} as indicated. Values of $T_0$ and $V(T_0)$ are listed in Table~\ref{table2}.
} 
\vspace{-4mm}
\label{volume}}
\end{center}
\end{figure}

\begin{figure}
\begin{center}
\includegraphics[width=0.95\linewidth]{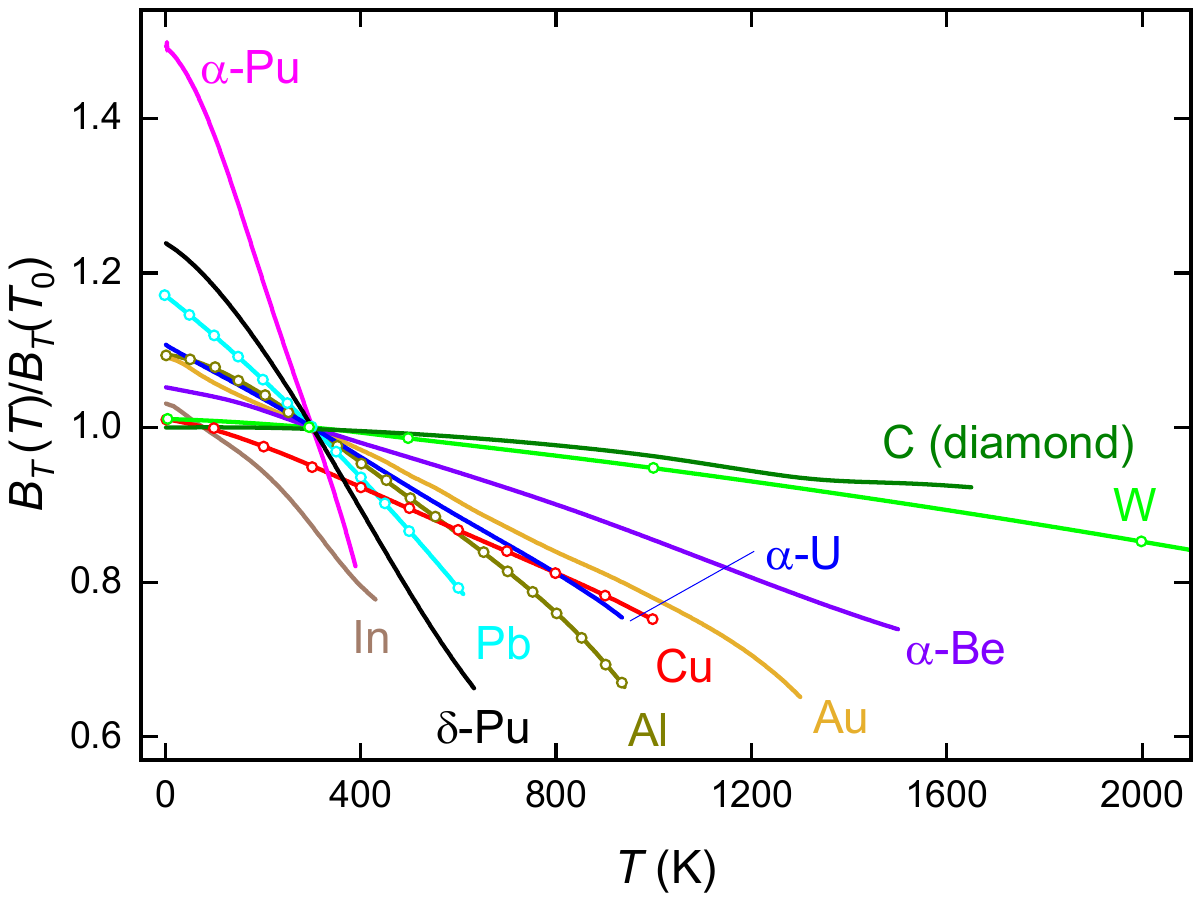}
\vspace{-4mm}
\textsf{\caption{Renormalized isothermal bulk modulus $B_T(T)/B_T(T_0)$ of $\alpha$-Be, Al~\cite{kozyrev2022}, Cu~\cite{kozyrev2023}, W~\cite{kozyrev2023i}, Pb~\cite{kozyrev2022i}, $\alpha$-U, $\alpha$-Pu and $\delta$-Pu as indicated. Values of $T_0$ and $B_T(T_0)$ are listed in Table~\ref{table2}. 
}
\vspace{-4mm}
\label{bulkiso}}
\end{center}
\end{figure}

In our study, Figures~\ref{volume}(a) and (b) respectively illustrate the normalized volume $V(T)/V(T_0)$ and thermal expansivity $\alpha_V(T)$, which are key parameters in calculating $C_{p,{\rm ph}}^{\rm QHA}(T)$ using the quasiharmonic approximation~\cite{jacobson2019, arblaster2018i, bodryakov2018, pamato2018, najashima2018, kozyrev2022, white1997, kozyrev2023, kozyrev2023i, kozyrev2022i, lloyd1966, harrison2023}. Additionally, Fig.~\ref{bulkiso} displays the temperature-dependent ratio of bulk modulus, $B_T(T)/B_T(T_0)$. For Al, Cu, W, and Pb, we have relied on experimental $B_T(T)$ values~\cite{kozyrev2022, kozyrev2023, kozyrev2023i, kozyrev2022i} that align with equation of state modeling.

\begin{figure}
\begin{center}
\includegraphics[width=0.95\linewidth]{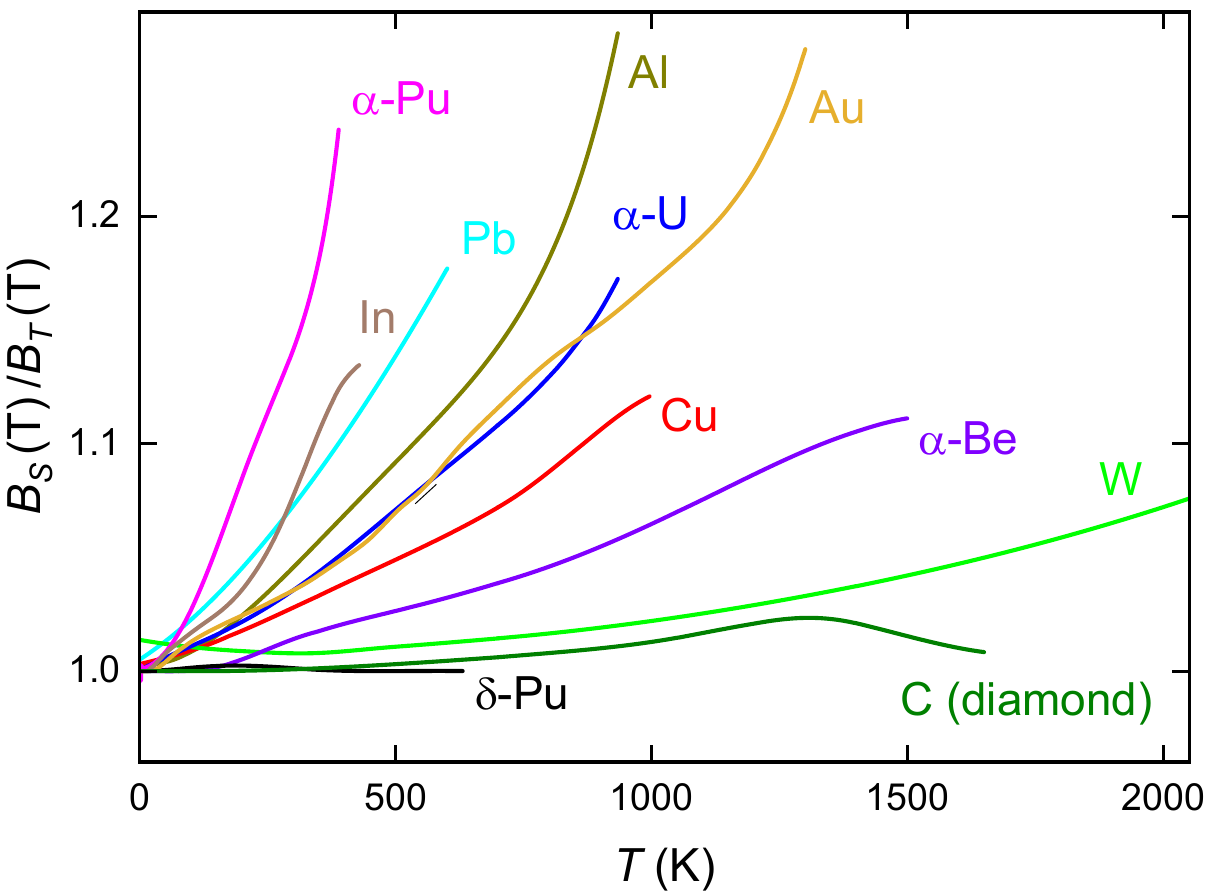}
\vspace{-4mm}
\textsf{\caption{Values of \textcolor{black}{$B_S(T)/B_T(T)$} obtained using Equation~(\ref{gamma2}).
}
\vspace{-4mm}
\label{gammafig}}
\end{center}
\end{figure}

For materials such as $\alpha$-Be, diamond, In, Au, $\alpha$-U, $\alpha$-Pu, and $\delta$-Pu, we calculated $B_T(T)$ using Equation~(\ref{gamma2}):
\textcolor{black}{\begin{equation}\label{gamma2}
\frac{B_S(T)}{B_T(T)}\equiv\frac{C_p(T)}{C_v(T)}  = 1 + \frac{V(T) \alpha_V^2(T) B_S(T) T}{C_p(T)}.
\end{equation}
This} calculation involves published values of $B_S(T)$ (shown in Fig.~\ref{shear}), $\alpha_V(T)$, and $C_p(T)$~\cite{bodryakov2018, lee1969, nakamura1980, gordon1976, oetting1983}. The resulting \textcolor{black}{$B_S(T)/B_T(T)$} values are graphically represented in Fig.~\ref{gammafig}.


\section*{APPENDIX B: LEAST SQUARES ANALYSIS METHOD}

\textcolor{black}{An examination of specific heat datasets reveals that the absolute value of the specific heat varies between different experimental curves, often in a manner that depends strongly on the temperature, whereas the individual points in each datasets are strongly correlated. In other words, some of the experimental curves are shifted relative to each other. For this reason, a traditional $\chi^2$ analysis, which treats individual points as being subject to equal uncertainty tends to underestimate the error. To account for the relative shift in the values of $C_p$ between curves, a summation was made of the form:}
\begin{equation}\label{leastsquares}
b(\eta) = \sum_i \big(C^{\rm ESA}_p(T_i) - C_p(T_i)\big)^2,
\end{equation}
\textcolor{black}{where} $i$ refers to individual specific heat data points $C_p(T_i)$ taken at a temperature $T_i$. 
For each elemental solid, the value of $\eta$ was estimated utilizing a least squares \textcolor{black}{minimization} analysis method, wherein $\eta$ is iteratively adjusted. 

To estimate the error bars \textcolor{black}{in $\eta$} for each elemental solid \textcolor{black}{arising from variations in the value of $C_p$ between curves (that include points that are strongly correlated)}, we assessed the change in $\eta$ necessary to double $b$. For elemental solids with multiple published specific heat datasets, \textcolor{black}{large} discrepancies between these datasets lead to larger uncertainties in determining $\eta$. \textcolor{black}{Note that individual curves were treated equally, including those that are already extracted from multiple datasets.}

\textcolor{black}{The estimated error in $\eta$ leads to an error in the anharmonic contribution $\sigma\Delta C_p^{\rm AN}$ (see Table~\ref{table2}). This value has been averaged over the temperature range. 
}


\section*{APPENDIX C: VALUES OF THE ADIABATIC BULK AND SHEAR MODULI}

In $\alpha$-U, $\alpha$-Pu and $\delta$-Pu, $B_S(T)$ and $G_S(T)$ were determined from measurements of polycrystalline samples~\cite{suzuki2011,armstrong1972}. In order to facilitate a direct comparison between different elemental solids with different crystalline structures, we have used $B_S(T)$ and $G_S(T)$ throughout. In the cases of diamond, Al, Cu, In, W, Au and Pb, $B_S(T)$ and $G_S(T)$ were estimated from elastic moduli measurements on single crystals. This was done by taking the geometric means of $B_S(T)$ and $G_S(T)$ estimated using the Voigt~\cite{voigt1928} and Reuss~\cite{reuss1929} methods, as suggested by Hill~\cite{hill1965}. 
Polynomials were used to fit $B_S(T)$ and $G_S(T)$. The parameters are listed in Table~\ref{table3}.

\begin{table*}
\begin{tabular}{||c|c|c|c|c|c|c|c|c||} 
 \hline
 element & $B_0$& $B_1$& $B_2$&$B_3$&$G_0$&$G_1$&$G_2$&$G_3$ \\ [0.5ex] 
 \hline\hline
$\alpha$-Be & 112.962&-0.0128315 & -2.45854$\times10^{-6}$ &0&149.726&-6.05829$\times10^{-5}$&-9.28125$\times10^{-6}$&0 \\ 
 \hline
diamond & 444.197& 0.00274012 &-1.4784$\times10^{-5}$ &1.06515$\times10^{-9}$&535.059&0.00266921&-2.14913$\times10^{-5}$&1.24698$\times10^{-9}$\\
 \hline
Al & 79.7224& -0.00663411 & -2.22926$\times10^{-5}$ &1.00555$\times10^{-8}$&29.6112&-0.00872952&-1.0797$\times10^{-5}$&4.58994$\times10^{-9}$\\
 \hline
Cu & 142.235& -0.00787221 & -3.93485$\times10^{-5}$ &2.56058$\times10^{-8}$&51.1213&-0.00831371&--2.55544$\times10^{-5}$&1.88028$\times10^{-8}$\\
 \hline
 In & 46.5552& -0.00947208 & -1.42592$\times10^{-5}$ &0& 7.83453&-0.0114104&0&0 \\
 \hline
W & 315.627& -0.0177789 & 3.43707$\times10^{-7}$ &0&163.69&-0.012698&-3.33073$\times10^{-6}$&0 \\
 \hline
Au & 181.641& -0.0358176 & 2.04844$\times10^{-5}$ &-1.44161$\times10^{-8}$&29.2705&-0.0038301&-5.81716$\times10^{-6}$&0 \\
 \hline
Pb & 48.9957& -0.0135392 & 4.78957$\times10^{-7}$ &-1.02712$\times10^{-8}$&11.2625&-0.00979439&0&0 \\ 
\hline
$\alpha$-U & 130.111& -0.0279822 & 0 &0&92.1994&-0.0200572&-3.79219$\times10^{-5}$&0 \\ 
\hline
$\alpha$-Pu & 72.8813& -0.0231165 & -0.000174498 &2.05688$\times10^{-7}$&59.4488 &-0.0200056&-0.000130367&1.53013$\times10^{-7}$\\ 
\hline
$\delta$-Pu & 37.8123& -0.0114931 & -5.62579$\times10^{-5}$ &4.81719$\times10^{-8}$ &20.2357&-0.00511441&-3.35896$\times10^{-5}$&2.70191$\times10^{-8}$ \\ 
 \hline
\end{tabular}
\caption{\label{table3} Polynomial fit parameters for $B_S(T)$ and $G_S(T)$, where $B_S(T)=B_0+B_1T+B_2T^2+B_3T^3$ and $G_S(T)=G_0+G_1T+G_2T^2+G_3T^3$. The units are in GPa.}
\end{table*}

In the case of Al, Cu, In, W and Pb, the elastic moduli were measured using an ultrasonic pulse-echo technique~\cite{kamm1964,gerlich1969, overton1955,chang1966, chandrasekhar1961,vold1977, neighbours1958,waldorf1962}. In the case of C (diamond), the elastic moduli below room temperature were measured using resonant ultrasound spectroscopy~\cite{migliori2008}, whereas the elastic moduli above room temperature were estimated using Brillouin scattering~\cite{zouboulis1998}. In $\alpha$-Be, $B_S(T)$ and $G_S(T)$ were estimated by means of high rate straining~\cite{dremov2009}. In $\alpha$-Pu and $\delta$-Pu, $B_S(T)$ and $G_S(T)$ were measured using resonant ultrasound spectroscopy~\cite{suzuki2011}. In $\alpha$-U, meanwhile, $B_S(T)$ and $G_S(T)$ were measured using a thin rod resonance method~\cite{armstrong1972}.

\section*{APPENDIX D: SPECIFIC HEAT RESIDUALS FOR THE ACTINIDE ELEMENTAL SOLIDS}

\begin{figure*}
\begin{center}
\includegraphics[width=0.60\linewidth]{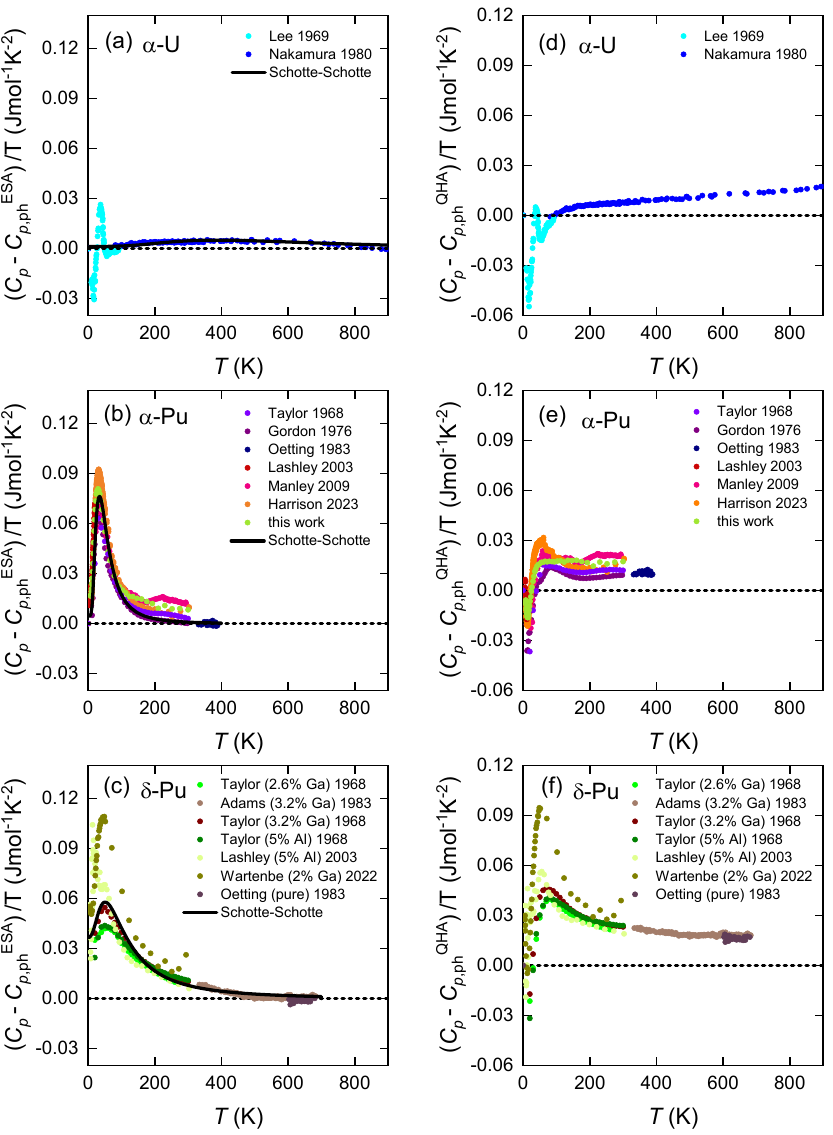} 
\vspace{-4mm}
\textsf{\caption{
The residual specific heat divided by temperature after having subtracted $C_{p,{\rm ph}}^{\rm ESA}(T)$ in $\alpha$-U (a), $\alpha$-Pu (b) and $\delta$-Pu (c), and after having subtracted $C_{p,{\rm ph}}^{\rm QHA}(T)$ in $\alpha$-U (d), $\alpha$-Pu (e) and $\delta$-Pu (f). Black solid lines in (a), (b) and (c) depict a Schotte-Schotte anomaly (discussed in Appendix E).  
}
\vspace{-4mm}
\label{actinideresiduals}}
\end{center}
\end{figure*}

Figure~\ref{actinideresiduals} illustrates that only through our elastic softening approximation method does the difference between the experimental $C_p(T)$ and the computed $C_{p,{\rm ph}}(T)$ yield a residual specific heat versus temperature, which aligns closely with a physically plausible non phonon contribution. In contrast, the quasiharmonic approximation tends to overestimate the phonon contribution to $C_p(T)$ at lower temperatures, leading to negative residuals for $\big(C_{p}(T)-C_{p,{\rm ph}}^{\rm QHA}(T)\big)/T$ in Pu. The observed peaks in $\big(C_{p}(T)-C_{p,{\rm ph}}^{\rm ESA}(T)\big)/T$ for $\alpha$-Pu and $\delta$-Pu could indicate several possibilities, such as a Schottky anomaly related to an energy gap or magnetic energy levels~\cite{harrison2019}, a Schotte-Schotte anomaly~\cite{schotte1975} from a resonance feature in the electronic density of states~\cite{harrison2023}, or the emergence of Kondo singlets~\cite{lawson2013}  (see discussion in Appendix E).


In the case of $\alpha$-U, considering $B_S(T)$ and $G_S(T)$ values only above 50 K, to circumvent discontinuities associated with charge-density wave phases, results in negative residuals at lower temperatures, even when employing our elastic softening approximation. Despite these limitations below approximately 50 K, the phase transition anomaly associated with the charge-density waves becomes distinctly evident in $\big(C_{p}(T)-C_{p,{\rm ph}}^{\rm ESA}(T)\big)/T$, as shown in Fig.~\ref{actinideresiduals}(a). 


\section*{APPENDIX E: CONTRIBUTIONS FROM NARROW BAND ELECTRONIC OR MAGNETIC DEGREES OF FREEDOM}

For illustrative purposes, we utilize the Schotte-Schotte anomaly model~\cite{schotte1975} to approximate possible contributions from magnetic or electronic degrees of freedom in actinide solids. This model may be relevant when considering excitations between different magnetic energy states or sharp features in the electronic density of states of an insulator or metal, offset relative to the chemical potential. If the energy levels are infinitely sharp, the result is a Schottky anomaly~\cite{tari2003}. In cases where these levels are broadened, creating a Lorentzian-shaped feature in the electronic density of states, the resultant specific heat feature aligns with the functional form of a Schotte-Schotte anomaly~\cite{schotte1975}. Originally proposed for $f$-electron compounds in strong magnetic fields, this model accounts for specific heat contributions in scenarios where a Kondo resonance undergoes Zeeman splitting~\cite{lacerda1989}.

The specific heat contribution of such an anomaly is described by 
\begin{equation}\label{schotteschotte}
C_{\rm el}(T)\approx\frac{R}{k_{\rm B}}\frac{\partial}{\partial T}\bigg[\int_{-\infty}^\infty\varepsilon D_{\rm el}(\varepsilon)f(\varepsilon/T){\rm d}\varepsilon\bigg],
\end{equation}
where \textcolor{black}{$f(\varepsilon/T)$ is the Fermi-Dirac distribution and}
\[D_{\rm el}(\varepsilon)=\frac{w\Gamma}{\pi((\varepsilon-\mu\pm\Delta)^2+\Gamma^2)}\] takes the form of a Lorentzian, \textcolor{black}{$\mu$ is the chemical potential}, and $\varepsilon$ is the energy. The spectral weight $w$ of the anomaly, and the parameters $\Delta$ and $\Gamma$, representing the energy gap and the width of the Lorentzian, respectively, define the anomaly's line shape. Specific heat curves resembling a Schotte-Schotte anomaly are common in valence fluctuating systems, often characterized by prominent, lifetime-broadened features in the electronic density of states~\cite{zhu2013,kushwaha2019}.

In $\alpha$-Pu, as shown in Fig.~\ref{actinideresiduals}(b), the consistency among different measurements allows us to approximate the residual with a Schotte-Schotte anomaly, characterized by parameters $\Delta=$~9.7~meV, $\Gamma=$~0.6~meV, and $w=$~0.46, similar to findings in Ref.~\cite{harrison2023}. For $\delta$-Pu, variations in the magnitude of the residuals, potentially influenced by Ga or Al concentration, are represented by a broader Schotte-Schotte anomaly, averaging $\Delta=$~17.4~meV, $\Gamma=$~10.2~meV, and $w=$~0.94. This may also reflect the possibility of a composition-dependent Schottky contribution at higher energies, as suggested in studies related to the invar effect~\cite{lawson2006,lawson2013}. In the case of $\alpha$-U at temperatures above 100 K, a broader anomaly with $\Delta=$~120~meV, $\Gamma=$~30~meV, and $w=$~0.6 is required to approximate the residual. However, this model becomes less representative at lower temperatures, particularly below 40~K, due to the charge-density wave phases in this system.

\section*{APPENDIX F: PREPARATION OF $\alpha$-Pu MATERIAL USED FOR MEASUREMENTS IN THIS WORK}
A chunk of $\alpha$-Pu was arc-melted in a multi-stage pumped/purged chamber employing a Zr-gettered argon atmosphere. The melting was performed to remove surface oxide present, to allow the $^4$He built up from radioactive decay to be released, and to remove defects formed from self-irradiation damage. Upon completion of arc-melting, the resulting clean button was wrapped into Ta-foil and then sealed within an evacuated fused silica tube. The resulting ampoule was then loaded into a box furnace, which was then ramped to 460~$^\circ$C, and the held there for one month. After one month, the ampoule was furnace cooled to retrieve the button.

\section*{APPENDIX G: $\alpha$-Pu SPECIFIC HEAT MEASUREMENTS PERFORMED IN THIS WORK}
Specific heat data using the fresh piece $\alpha$-Pu button was collected using a 14-Tesla Quantum Design DynaCool Physical Property Measurement System (PPMS). The semi-adiabatic pulse technique was employed. Contributions of Apiezon N-grease used to mount the sample to the sapphire stage of the Quantum Design specific heat option were subtracted using a separate measurement of the addenda.

\section*{\textcolor{black}{APPENDIX H: MODELS CLAIMING NEGLIGIBLE ANHARMONICITY IN Pu}}
\textcolor{black}{To bolster their theoretical claim of negligible anharmonicity in $\delta$-Pu, Ref.~\onlinecite{soderlind2023} compared their phonon-specific-heat calculations with an estimate from Lashley \emph{et al.}~\cite{lashley2003}. 
Specifically, Lashley \emph{et al.}\ generated discrete phonon heat-capacity curves from the phonon density of states measured at a few fixed temperatures~\cite{mcqueeney2004}, then interpolated among these discrete curves to produce a continuous one.}

\textcolor{black}{However, by interpolating among these discrete {specific-heat} curves {rather than} using an {entropy}-based approach (as in Equation~\ref{phononenytropy}), Lashley \emph{et al.}\ effectively substituted shifted phonon frequencies (\emph{i.e.}, phonon softening) {into the specific heat rather than the entropy}, a procedure known to give incorrect results~\cite{barron1965,wallace1972,jacobs2005,allen2015}. In other words, by excluding derivatives of the form $\partial \omega / \partial T$, Ref.~\onlinecite{lashley2003} inadvertently omitted anharmonic contributions to the entropy from their phonon heat-capacity estimate. Indeed, Ref.~\onlinecite{lashley2003} explicitly states that anharmonic effects were neglected under the assumption that $\partial \omega / \partial T$ is small; in fact, $\partial \omega / \partial T$ is anomalously large in $\delta$-Pu~\cite{mcqueeney2004}.}


%

\section{Data availability}
The data that support the findings of this study are available upon reasonable request.

\bibliographystyle{naturemag}

\section{Acknowledgements}
Support was provided by Los Alamos National Laboratory 
LDRD project 20230042DR (code XXN0). 
A portion of this work was carried out at the National High Magnetic Field Laboratory, 
which is funded by NSF Cooperative Agreement 1164477, the State of Florida and DoE. DOE BES project ``Science of 100 tesla'' supported the modeling of electronic contribution. 

\end{document}